\documentclass{lmcs}

\usepackage{etoolbox}

\usepackage[T1]{fontenc}
\usepackage[utf8]{inputenc}
\usepackage[%
]{microtype}
\usepackage[american]{babel} 

\usepackage{etoolbox}

\usepackage{bm} 

\usepackage{stmaryrd}
\usepackage{graphicx}
\usepackage{xspace,xcolor}
\usepackage{amsmath,amssymb,mathtools} 

\usepackage{rsfso}
\usepackage{scalerel}
\usepackage{mathpartir}
\usepackage{prftree}
\usepackage{bussproofs}
\usepackage{lineno}
\usepackage{marvosym}

\usepackage{ellipsis}
\usepackage{placeins} 

\usepackage{tikz}
\usepackage{tikz-cd}
\usepackage{pgfplots}

\usepgfplotslibrary{fillbetween}
\usetikzlibrary{intersections}
\pgfplotsset{compat=1.14}

\usetikzlibrary{arrows}
\usetikzlibrary{arrows.meta}
\usetikzlibrary{decorations.pathmorphing,shapes}

\pgfdeclarelayer{bg}
\pgfsetlayers{bg,main}
\usetikzlibrary{decorations.pathreplacing,decorations.markings}

\definecolor{pamblue}{rgb}{.78, .90, .98}

\newcommand{\kTileHeight}{.5em}

\newtheorem*{theorem*}{Theorem}

\newcommand{\iwouldprefernotooverline}[1]{#1}

\newcommand{\s}{\sigma}
\newcommand{\g}{\gamma}

\newcommand{\state}{\sigma}
\newcommand{\mstate}{\mathfrak{s}}
\newcommand{\memstate}{\mu}
\newcommand{\statevector}{\boldsymbol{\sigma}}
\newcommand{\statevectorstar}{\boldsymbol{\sigma}^{*}}

\DeclarePairedDelimiter\sem{\llbracket}{\rrbracket}
\DeclarePairedDelimiter\abs{\lvert{}}{\rvert}

\newcommand{\finalnodes}[1]{\abs{#1}}

\newcommand{\alter}[3]{#1\{#2\mapsto #3\}}  

\newcommand{\readarea}[1]{\textrm{rd}(#1)}
\newcommand{\writearea}[1]{\textrm{wr}(#1)}
\newcommand{\allocarea}[1]{\textrm{mem}(#1)}

\newcommand{\edgesource}[0]{\partial^{-}}
\newcommand{\edgetarget}[0]{\partial^{+}}
\newcommand{\sourcefn}{\partial_0}

\newcommand{\targetfn}{\partial_1}

\newcommand{\State}{\kl[memory state]{\mathbf{State}}}
\newcommand{\MStates}{\kl[machine state]{\mathbf{MState}}}
\newcommand{\SStates}{\kl[separated state]{\mathbf{SState}}}

\newcommand{\LogStates}{\kl[logical state]{{\mathbf{LState}}}}
\newcommand{\LStates}{\LogStates}
\newcommand{\Code}{\mathbf{Code}}
\newcommand{\Instr}{\kl[instruction]{\mathbf{Instr}}}
\newcommand{\RVar}{\kl[resource]{\mathbf{LockName}}}
\newcommand{\Var}{\kl[variable names]{\mathbf{Var}}}
\newcommand{\Loc}{\kl[location]{\mathbf{Loc}}}
\newcommand{\Val}{\mathbf{Val}}

\newcommand{\sequential}{;}

\newcommand{\Perm}{\kl[permission]{\mathbf{Perm}}}

\newcommand{\bool}{\mathbf{bool}}

\DeclareMathOperator{\dom}{dom}
\DeclareMathOperator{\domC}{dom_C}
\DeclareMathOperator{\domF}{dom_F}

\DeclareMathOperator{\hdom}{hdom}
\DeclareMathOperator{\vdom}{vdom}

\DeclareMathOperator{\lock}{lock}
\DeclareMathOperator{\True}{\mathbf{true}}
\DeclareMathOperator{\False}{\mathbf{false}}

\newcommand{\kw}[1]{\mathbb{\mathtt{#1}}}
\newcommand{\kwhile}{\kw{while}}
\newcommand{\kif}{\kw{if}}
\newcommand{\kthen}{\kw{then}}
\newcommand{\kelse}{\kw{else}}
\newcommand{\kalloc}{\kw{alloc}}
\newcommand{\kmalloc}{\kw{malloc}}
\newcommand{\kdispose}{\kw{dispose}}
\newcommand{\kwith}{\kw{with}}
\newcommand{\kwhen}{\kw{when}}
\newcommand{\kresource}{\kw{resource}}
\newcommand{\knop}{\kw{nop}}
\newcommand{\kskip}{\kw{skip}}
\newcommand{\kdo}{\kw{do}}
\newcommand{\ifte}[3]{\kif\: #1\: \kthen\: #2\: \kelse\: #3}
\newcommand{\while}[2]{\kwhile\: #1 \: \kdo \: #2}
\newcommand{\resource}[2]{\kresource\: #1 \: \kdo \: #2}
\newcommand{\when}[3]{\kwith\: #1 \: \kwhen\: #2 \: \kdo \: #3}

\newcommand{\deref}[1]{[#1]}

\newcommand{\powerset}{\raisebox{.15\baselineskip}{\Large\ensuremath{\wp}}}

\DeclareMathOperator*{\bigsprod}{\scalerel*{\circledast}{\sum}}

\newcommand{\PAR}{\text{\sc Par}}
\newcommand{\IF}{\text{\sc If}}

\newcommand{\CONJ}{\text{\sc Conj}}
\newcommand{\DISJ}{\text{\sc Disj}}
\newcommand{\WHEN}{\text{\sc With}}
\newcommand{\RES}{\text{\sc Res}}
\newcommand{\AFF}{\text{\sc Aff}}
\newcommand{\SEQ}{\text{\sc Seq}}
\newcommand{\FRAME}{\text{\sc Frame}}

\newcommand{\STORE}{\text{\sc Store}}
\newcommand{\LOAD}{\text{\sc Load}}

\newcommand{\fv}{\mathrm{fv}}

\newcommand{\defpred}{\mathrm{def}}

\DeclareMathOperator{\own}{Own}

\makeatletter

\newbox\prf@@fancysummarybox
\newdimen\prf@@fancysymmarylen
\def\prffancysummarybox{%
  \sbox{\prf@@fancysummarybox}{$\vdots$}%
  \prf@@fancysymmarylen\ht\prf@@fancysummarybox%
  \advance\prf@@fancysymmarylen\dp\prf@@fancysummarybox%
  \sbox{\prf@@fancysummarybox}{%
    \raisebox{.25\prf@@fancysymmarylen}[.8\prf@@fancysymmarylen]%
    [0pt]{\usebox{\prf@@fancysummarybox}}}%
  \prf@@fancysymmarylen\wd\prf@summary@label%
  \ifdim\prf@@fancysymmarylen>\z@\relax%
    \prf@@fancysymmarylen\wd\prf@@fancysummarybox%
    \wd\prf@summary@label.4em%
    \hbox to\prf@@fancysymmarylen{%
      \usebox\prf@@fancysummarybox}\kern-.4em%
      \box\prf@summary@label%
  \else\usebox\prf@@fancysummarybox\fi}

\newcommand{\oset}[3][0ex]{%
  \mathrel{\mathop{#3}\limits^{
    \vbox to#1{\kern-2\ex@
    \hbox{$\scriptstyle#2$}\vss}}}}
\makeatother

\newcommand{\raiselabel}[1]{\raisebox{0.4em}{#1}}

\newcommand{\pointsto}[1]{\oset{\text{\tiny $#1$}}{\mapsto}}

\newcommand{\emp}{\mathbf{emp}}

\newcommand{\triple}[4]{#1 \!\vdash\!\{#2\} #3 \{#4\}}
\newcommand{\triplestd}{\triple{\Gamma}{P}{C}{Q}}

\newcommand{\parfinmap}{\rightharpoonup_{\mathit{fin}}}

\newcommand{\updmap}[3]{#1[#2 \mapsto #3]}

\newcommand\evalexpr[3]{#1(#2) = #3}

\newcommand{\kArrowLength}{1cm}

\newcommand{\redto}[3]{%
    \begin{tikzcd}[ampersand replacement=\&, column sep=\kArrowLength]%
      {#1} \arrow[r, rightsquigarrow, "#2"] \& {#3}%
    \end{tikzcd}%
  }

  \newcommand{\yrightarrow}[1]{\,\xrightarrow{\;\;\;#1\;\;\;}\,}

\newcommand{\hide}[1]{\mathsf{hide}[#1]}

\newcommand{\whentrue}[1]{\mathsf{whentrue}[#1]}
\newcommand{\whenfalse}[1]{\mathsf{whenfalse}[#1]}
\newcommand{\whenabort}[1]{\mathsf{whenabort}[#1]}

\newcommand{\kparproj}{\mathbf{proj}}

\newcommand{\parprojl}{\lambda_1}
\newcommand{\parprojr}{\lambda_2}

\newcommand{\transitionpath}{\twoheadrightarrow}

\newcommand{\semS}[1]{\sem{#1}_S}
\newcommand{\semSclosed}[2]{\sem{#1}_{S}^{#2}}
\newcommand{\semL}[1]{\sem{#1}_L}
\newcommand{\semSep}[1]{\sem{#1}_{\mathit{Sep}}}
\newcommand{\Sep}{Sep}

\newcommand{\juxt}{\oplus}
\newcommand{\getstate}[1]{\lambda_{#1}}
\newcommand{\parprod}{\parallel}
\newcommand{\high}[1]{#1}

\newcommand{\fp}{\rho}

\newcommand{\footprint}{\mathrm{footprint}}
\newcommand{\interstate}[2]{#1 \&\: #2}

\newcommand{\Error}{\lightning}

\newcommand{\framing}[1]{\mathrm{frame}[#1]}
\newcommand{\ATSwhen}[1]{\mathrm{when}[#1]}

\newcommand{\take}[1]{\mathrm{acquire}[#1]}
\newcommand{\release}[1]{\mathrm{release}[#1]}

\newcommand{\SepModelGraph}{\kl[machine model of separated states]{\genFPgraph_{\mathit{Sep}}}}
\newcommand{\StateModelGraph}{\kl[stateful model]{\genFPgraph_S}}
\newcommand{\LockModelGraph}{\kl[stateless model]{\genFPgraph_L}}

\newcommand{\semmorphfull}{\mathcal{L}}    
\newcommand{\semmorphsep}{\mathcal{S}}      
\newcommand{\semmorph}{\mathcal{L}}     

\newcommand{\Locks}{\kl[well-defined locks]{\mathit{Locks}}}

\newcommand{\ATS}{\mathbf{ATS}}

\newcommand{\genFPgraph}{\text{\Large $\moo$}}

\newcommand{\valuev}{v}
\newcommand{\location}{\ell}

\newcommand{\CODE}[1]{{#1}\, : \,C}
\newcommand{\ENV}[1]{#1 \, : \,F}
\newcommand{\ENVbig}[1]{#1:F}
\newcommand{\CODEbig}[1]{#1:C}

\def\bendamnt{13}

  \tikzset{
    vertex/.style={text centered, fill, color=black, circle, inner sep=.7pt},
    state/.style={},
    every label/.style={label distance=-2pt},
    every label/.append style={font=\footnotesize},
    labelle/.style={%
      postaction={ decorate,
        decoration={ markings, mark=at position .5 with \node #1;}}}}

\newcommand{\squarediag}[8]{
\begin{tikzpicture}[scale=0.85,]

  \node[vertex, draw] (a) at (-1,0) [label=left:$#1$] {};
  \node[vertex, draw] (d) at (1,0) [label=right:$#2$] {};
  \node[vertex, draw] (b) at (0,1) [label=above:$#3$] {};
  \node[vertex, draw] (c) at (0,-1) [label=below:$#4$] {};

  \begin{pgfonlayer}{bg}
    \fill[pamblue] (a.center) to [bend left=\bendamnt] (b.center)
                   to [bend left=\bendamnt] (d.center)
                   to [bend left=\bendamnt] (c.center)
                   to [bend left=\bendamnt] (a.center);
  \end{pgfonlayer}
  \begin{scope}[line width=.45pt,decoration={
       markings,
       mark=at position 0.56 with {\arrow{Stealth}}}
     ] 
     \draw[postaction={decorate}, labelle={[above left]{\footnotesize $#5$}}] (a.center) to [bend left=\bendamnt] (b.center);
     \draw[postaction={decorate}, labelle={[above right]{\footnotesize $#6$}}] (b.center) to [bend left=\bendamnt] (d.center);
     \draw[postaction={decorate}, labelle={[below left]{\footnotesize $#7$}}] (a.center) to [bend right=\bendamnt] (c.center);
     \draw[postaction={decorate}, labelle={[below right]{\footnotesize $#8$}}] (c.center) to [bend right=\bendamnt] (d.center);
   \end{scope}
\end{tikzpicture}
}

\graphicspath{{dessins/rewriting/}}

\def\paragraph{\subsubsection*}

\usepackage{todonotes}

\usepackage{expl3}
\ExplSyntaxOn
\let\ior_get_str:NN\ior_str_get:NN
\ExplSyntaxOff

\usepackage{hyperref}
\usepackage[xcolor,quotation,notion,electronic]{knowledge}

\knowledgeconfigure{protect quotation={tikzcd}}

\knowledge{asynchronous transition system}[asynchronous transition systems|ATS|ATSs|Asynchronous transition system|Asynchronous transition systems]{notion}

\knowledge{returning node}[returning nodes|returning states|returning state]{notion}

\knowledge{initial}[initial nodes|initial node|initial state|initial states]{notion}
\knowledge{final}[final nodes|final node|final state|final states]{notion}

\knowledge{Asynchronous Machine Models}[machine model|machine models|Machine Model]{notion}

\knowledge {asynchronous graph homomorphism}[asynchronous graph morphism|asynchronous graph morphisms|asynchronous morphism|asynchronous morphisms|morphism of asynchronous graphs]{notion}
\knowledge {asynchronous graph}{notion}
\knowledge {permutation tile}[tile|tiles|permutation tiles]{notion}
\knowledge {equivalent modulo one permutation tile}[equivalent|homotopy]{notion}
\knowledge {resource}[resources|lock|locks|resource variable|resource variables]{notion}
\knowledge {machine instruction}[machine instructions|instruction|instructions]{notion}
\knowledge {variable names}[variable|variables]{notion}
\knowledge {memory location}[memory locations|location|locations]{notion}
\knowledge {Memory states}[memory state|memory states]{notion}
\knowledge {Machine states}[machine states|machine state]{notion}
\knowledge {separated state}[separated states]{notion}
\knowledge {machine step}{notion}
\knowledge {expression}{notion}
\knowledge {machine model of separated states}[The machine model of separated states|Machine model of Separated States]{notion}
\knowledge {stateful model}{notion}
\knowledge {stateless model}{notion}
\knowledge {machine state footprint}{notion}
\knowledge {lock footprint}{notion}
\knowledge {well-defined locks}{notion}
\knowledge {logical state}[logical states]{notion}
\knowledge {permissions}[permission|permission monoid]{notion}
\knowledge {predicate}[predicates|formulas|formula]{notion}
\knowledge {invariant}[invariants]{notion}
\knowledge {precision}[precise]{notion}
\knowledge {parallel product}{notion}
\knowledge {sequential product}[sequential composition]{notion}
\knowledge {proof tree}[derivation tree|derivation trees|proof]{notion}
\knowledge {$1$-fibration}[$1$-fibrations]{notion}
\knowledge {Hoare triple}[Hoare triples]{notion}
\knowledge {2-fibration}[2-fibrations|$2$-fibration|$2$-fibration|$2$-fibrational]{notion}
\knowledge {natural}{notion}
\knowledge {artificial}{notion}
\knowledge {footprints}[footprint]{notion}
\knowledge {Code $1$-fibration}[$1$-fibration on Code transitions]{}

\newtoggle{enable comments}

\toggletrue{enable comments}

\iftoggle{enable comments}{
  \newcommand{\ls}[1]{\todo[inline, caption={2do},color=lime]{\hspace{.9em}\begin{minipage}{\textwidth}LS: #1\end{minipage}}}
  \newcommand{\pam}[1]{\todo[inline]{LS: #1}}
}{
  \newcommand{\ls}[1]{}
  \newcommand{\pam}[1]{}
}

\begin{document}


\title[An Asynchronous Soundness Theorem for CSL]{An Asynchronous Soundness Theorem\\ for Concurrent Separation Logic}

\author{Paul-Andr\'e Melli\`es}
\author{L\'eo Stefanesco}

\address{Institut de Recherche en Informatique Fondamentale (IRIF), CNRS \& Université
  Paris Diderot}

\begin{abstract}
  Concurrent separation logic (CSL) is a specification logic for concurrent
  imperative programs with shared memory and locks.
  In this paper, we develop a concurrent and interactive account of the logic
  inspired by asynchronous game semantics.
  To every program~$C$, we associate a pair of asynchronous transition systems
  $\semS{C}$ and $\semL{C}$ which describe the operational behavior of the Code
  when confronted to its Environment or Frame --- both at the level of machine
  states~($S$)
  and of machine instructions and locks~($L$).
%
  We then establish that every derivation tree~$\pi$ of a judgment
  $\Gamma\vdash\{P\}C\{Q\}$ defines a winning and asynchronous
  strategy~$\semSep{\pi}$ with respect to both asynchronous semantics~$\semS{C}$
  and~$\semL{C}$.
%
%
%
  From this, we deduce an asynchronous soundness theorem for CSL, which states
  that the canonical map $\mathcal{L}:\semS{C}\to\semL{C}$ from the stateful
  semantics~$\semS{C}$ to the stateless semantics~$\semL{C}$ satisfies a basic
  fibrational property.
%
  We advocate that this
  provides a clean and conceptual explanation for the usual soundness theorem of
  CSL, including the absence of data races.
\end{abstract}

\maketitle

\section{Introduction}
A simple way to understand an imperative (possibly nondeterministic) program~$C$
is to interpret it as a binary relation $[C]\subseteq S\times S$ between machine
states $s,s'\in S$.
In that approach, the statement $s[C]s'$ indicates that one execution trace (at
least) of the program $C$ has initial state $s\in S$ and final state $s'\in S$.
One practical advantage
of this description is that the binary relation $[C]$ abstracts away from
the execution traces of the program~$C$, and only retains their initial and
final states.
However crude, this abstraction is generally sufficient to analyze the
properties of sequential imperative programs, and to establish the soundness of
Hoare logic.
Unfortunately, the abstraction becomes too coarse when one decides to shift to
concurrent imperative programs with shared memory and locks, and to establish
the soundness of a specification logic like Concurrent Separation Logic (CSL).
To that purpose, it has long been recognized that one needs a proper account of
the execution traces of the program~$C$, see Brookes~\cite{Brookes:a-semantics}.
In this paper, we go one step further, and advocate that the soundness theorem
of CSL, and more specifically the absence of data races, is intrinsically
related to the asynchronous structure of the execution paths of~$C$.
Inspired by asynchronous game semantics, we interpret every concurrent
imperative program~$C$ as a pair of asynchronous graphs~$\semS{C}$
and~$\semL{C}$ related by an "asynchronous graph homomorphism"
\begin{equation}\label{equation/lockmap}
\begin{tikzcd}[column sep=1.5em]
\mathcal{L}_C & : & 
\semS{C} \arrow[rr]
&&
\semL{C}
\end{tikzcd}
\end{equation}
We thus start by recalling the notion of \emph{"asynchronous
  graph"}~\cite{PAM:async-graph,mellies-hdr} before discussing the relationship
between time and space separation.

\medbreak

\paragraph{Asynchronous graphs} 
A graph $G=(V,E,\edgesource, \edgetarget)$ consists of a set~$V$ of vertices or
nodes, a set of $E$ of edges or transitions, and a source and a target function
$\edgesource,\edgetarget:E\to V$.
An ""asynchronous graph"" $(G,\diamond)$ is a graph~$G$ equipped with a binary
relation~$\,\diamond\,$ between paths $f,g: P\twoheadrightarrow Q$ of length 2,
with the same source and target vertices.
%
%
A pair $(f,g)$ such that $f\diamond g$ is called a \emph{permutation tile} and
is depicted as a 2-dimensional tile between the paths $f=u\cdot v'$ and
$g=v\cdot u'$ as follows:
\begin{equation}\label{equation/permutation-tile}
\raisebox{-2.8em}{\includegraphics[height=5.5em]{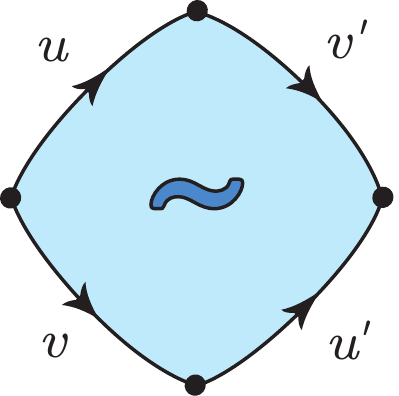}}
\end{equation}
The intuition conveyed by such a permutation tile $u\cdot v'\diamond v\cdot u'$
is that the two transitions~$u$ and~$v$ are independent.
For that reason, the two paths~$u\cdot v'$ and~$v\cdot u'$ may be seen as
equivalent up to scheduling.
%
%
%
%
The binary relation~$\,\diamond\,$ is required to satisfy the following two
axioms axiomatic properties below.

\noindent 
\emph{\textbf{Axiom 1.}} The permutation relation $\diamond$ is symmetric, in
the sense that $u\cdot v' \,\diamond\, v\cdot u'$ implies $v\cdot u'
\,\diamond\, u\cdot v'$ for all transitions~$u$, $v$,~$u'$,~$v'$.


\noindent 
\emph{\textbf{Axiom 2.}} In the situation below where $u\cdot w_1\diamond
v_1\cdot u_1$ and $u\cdot w_2\diamond v_2\cdot u_2$, one has that $v_1=v_2$ if
and only if $w_1=w_2$.
\[
\begin{array}{ccc}
  \includegraphics[width=5.5em]{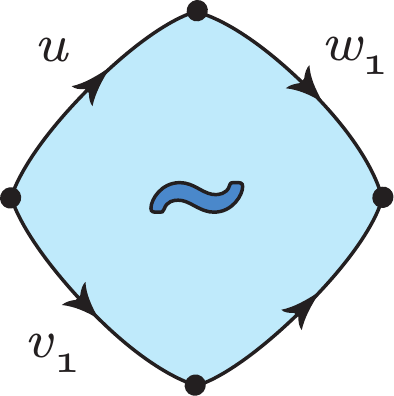}
  & \quad\quad\quad
  &
    \includegraphics[width=5.5em]{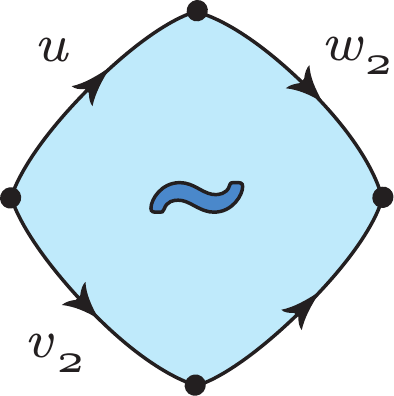}
\end{array}
\]

Two paths $f,g:M\transitionpath N$ of an "asynchronous graph" are
\emph{""equivalent modulo one permutation tile""} $h_1\diamond h_2$ when $f$ and $g$
factor as $f=d\cdot h_1\cdot e$ and $g=d\cdot h_2\cdot e$ for two paths
$d:M\transitionpath P$ and $e:Q\transitionpath N$.
%
We write $f\sim g$ when the path~$f:M\transitionpath N$ is "equivalent" to the
path $g:M\transitionpath N$ modulo a number of such permutation tiles.
Note that the relation $\sim$ is symmetric, reflexive and transitive, and thus
defines an equivalence relation, closed under composition.
%

\paragraph{Separation in space and time}
The 2-dimensional permutation tiles $f\diamond g$
provide a topological means to reflect the \emph{temporal} nature of
independence in concurrency theory.
Every permutation tile~(\ref{equation/permutation-tile}) indicates that the two
transitions~$u$ and~$v$ are independent in time: they may be equivalently
executed in the sequential order $u\cdot v'$ or in the sequential order $v\cdot
u'$.
Although all the asynchronous graphs considered in this paper are discrete, it
is enlightening to take the topological intuition of ``homotopy'' seriously, and
to imagine that the path $u\cdot v'$ could be transformed ``continuously'' into
the path $v\cdot u'$ by a sequence of local deformations of the form
\[
\includegraphics[width=5.5em]{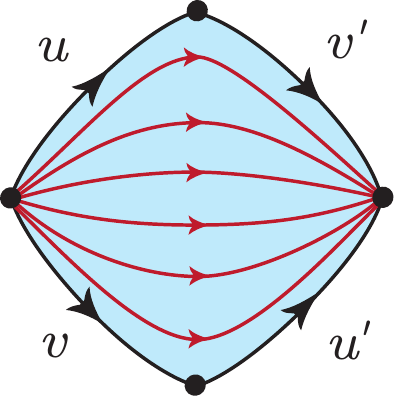}
\]
as it would be possible if one embedded our asynchronous graphs $(G,\diamond)$
in the topological framework of directed homotopy, see \cite{dihomotopy-book}.
In the same spirit, we could replace our 2-dimensional
graphs by higher-dimensional automata admitting $n$-dimensional
cubes~\cite{HDA}.

\medbreak

Interestingly, in practical situations, the \emph{temporal} independence of two
transitions~$u$ and~$v$ is not primitive: it is a consequence of their
\emph{spatial} separation.
In that respect, the idea of temporal independence may be seen as a layer of
abstraction above the more concrete and machine-dependent idea of spatial
separation.
We illustrate this basic but important point by constructing an asynchronous
graph $(G,\diamond_G)$ based on a very simple machine model,
consisting of
\begin{itemize}
\item a countable set $\Var$ of variables, written $x$, $y$, \textellipsis,
\item a countable set~$\Val$ of values, written $v$, $w$, \textellipsis,
\item a countable set~$\Loc\subseteq\Val$ of memory locations, written
  $\location$.
\end{itemize}
A \emph{memory state} $\memstate=(s,h)$ of the machine is defined as a pair
consisting of two partial functions
\begin{equation}\label{equation/basic-memory-states}
s: \Var \parfinmap \Val
\quad\quad\quad
h: \Loc \parfinmap \Val
\end{equation}
with finite domains, called the \emph{stack}~$s$ and the \emph{heap}~$h$ of the
memory state~$\memstate$.
%
%
The instructions~$m$ of our machine are of three kinds: 
\begin{equation}\label{equation/three-instructions}
x:=\valuev \quad\quad\quad
x:=[\location] \quad\quad\quad
[\location]:=x
\end{equation}
where (1) the instruction $x:= \valuev$ assigns a value $\valuev$ to the
variable~$x$, (2) the instruction $x:=[\location]$ loads the value
$h(\location)$ at location $\location$ and assigns it to the variable~$x$, and
(3) the instruction $[\location]:=x$ stores at location~$\location$ the current
value $s(x)$ of the variable~$x$.
The asynchronous graph $(G,\diamond_G)$ is defined in the following way.
Its nodes are the memory states~(\ref{equation/basic-memory-states}) of the
machine, and its transitions are of the form
%
\[
\begin{tabular}{cl}
\begin{tikzcd}[column sep=1.5em]
(s,h)
\ar[rr,"{x:=\valuev}"]
& {} &
(s',h)
\end{tikzcd}
&
when $s'=\alter{s}{x}{\valuev}$,
\\
$\begin{tikzcd}[column sep=1.5em]
(s,h)
\ar[rr,"{x:=[\location]}"]
& {} &
(s',h)
\end{tikzcd}$
&
when $h(\location)$ is defined and\\
& $s'=\alter{s}{x}{h(\location)}$,
\\
$\begin{tikzcd}[column sep=1.5em]
(s,h)
\ar[rr,"{[\location]:=x}"]
& {} &
(s,h')
\end{tikzcd}$
&
when $s(x)$ is defined and
\\
&
$h'=\alter{h}{\location}{s(x)}$.
\end{tabular}
\]
Here, we use the following convenient notation:
given a partial function $f:X\parfinmap Y$ with finite domain between two sets
$X$ and $Y$, and an element $y\in Y$, we write $\alter{f}{x}{y} : X\parfinmap Y$
for the partial function with finite domain defined as
\[
\alter{f}{x}{y} \quad : \quad 
x' \quad \mapsto \quad \left\{
\begin{array}{cl}
f(x) & \mbox{when $x'\neq x$,}
\\
y  & \mbox{when $x'= x$.}
\end{array}\right.
\]
In order to define the permutation tiles of the asynchronous graph
$(G,\diamond_G)$, one observes that every transition
\[\begin{tikzcd}[column sep=3em]
u \quad : \quad (s,h)
\ar[rr,"{m}"]
& {} &
(s',h')
\end{tikzcd}\]
performed by an instruction~$m$ reads and writes on a specific area
\[
\readarea{u}\subseteq \Var+\Loc
\quad\quad\quad
\writearea{u}\subseteq \Var+\Loc
\]
of the memory of the machine, which we shall call its \emph{footprint}.
This footprint may be computed from the instruction~$m$ performing the
transition $u=(\memstate,m,\memstate')$ in the following way:
\begin{center}
\fbox{$
\begin{array}{c}
\begin{array}{ccc}
\readarea{x:=\valuev}
&\!\!\! = \!\!\!& 
\emptyset
\\
\writearea{x:=\valuev}
& \!\!\! = \!\!\! & 
\{x\}
\end{array}
\\
\\
\begin{array}{ccccccc}
\readarea{x:=[\location]} & \!\!\! = \!\!\! & \{\location\}
& &
\readarea{[\location]:=x} & \!\!\! = \!\!\! & \{x\}
\\
\writearea{x:=[\location]} & \!\!\! = \!\!\!& \{x\}
  & &
\writearea{[\location]:=x} & \!\!\! = \!\!\! & \{\location\}
\end{array}
\end{array}$}
\end{center}
%
Now, suppose given two transitions $u:\memstate\to \memstate_1$ and
$v:\memstate\to \memstate_2$ starting from the same memory state $\memstate$ in
the graph~$G$.
The two transitions $u$ and $v$ are declared \emph{independent} when
\[
(\readarea{u}\cup\writearea{u}) \cap \writearea{v}=\emptyset
\quad \mbox{and} \quad
\writearea{u}\cap (\readarea{v}\cup\writearea{v})=\emptyset.
\]
Note that the independence of the transitions $u$ and $v$ is a
\emph{consequence} of their spatial separation.
It is not difficult to see that 
for every pair of such independent transitions
\[
\begin{array}{ccc}
\begin{tikzcd}[column sep=2em]
u \,\, : \,\, \mu_1
\ar[rr,"{m_1}"]
&&
\mu_2\end{tikzcd}
&\quad&
\begin{tikzcd}[column sep=2em]
v \,\, : \,\, \mu_2
\ar[rr,"{m_2}"]
&&
\mu_3\end{tikzcd}
\end{array}
\]
there exists a unique memory state $\mu_2'$ such that 
\[
\begin{array}{ccc}
\begin{tikzcd}[column sep=2em]
u' \,\, : \,\, \mu_2'
\ar[rr,"{m_1}"]
&&
\mu_3\end{tikzcd}
&\quad&
\begin{tikzcd}[column sep=2em]
v' \,\, : \,\, \mu_1
\ar[rr,"{m_2}"]
&&
\mu_2'\end{tikzcd}
\end{array}
\]
are transitions of the graph~$G$.
In that case, we say that $u'$ is the residual of $u$ after~$v$ and,
symmetrically, that $v'$ is the residual of $v$ after~$u$.
%
This basic confluence property leads us to the following definition.
%
A permutation tile of the form~(\ref{equation/permutation-tile}) 
\[
u\cdot v' \quad \diamond_G \quad v\cdot u'
\]
in the "asynchronous graph"~$(G,\diamond_G)$ is defined as a pair of independent
transitions~$u$ and~$v$ where the transition $u'$ is defined as the residual of
$u$ after~$v$, and the transition~$v'$ is defined as the residual of $v$
after~$u$.
It is not difficult to see that the graph $G=(V,E)$ of memory states and
transitions between them, together with the notion of permutation tile $u\cdot
v'\diamond_G v\cdot u'$ just defined, satisfy the axioms required of an
asynchronous graph $(G,\diamond_G)$.

%
%
\medbreak

\paragraph{Stateful vs. stateless semantics}
%
Along the \emph{stateful} description of the machine provided by the
asynchronous graph $(G,\diamond_G)$, comes a \emph{stateless} description of the
same machine, conveyed this time by an asynchronous graph~$(H,\diamond_H)$ where
only the instructions are considered, not their action on the machine states.
%
%
Accordingly, the graph~$H$ has a single node~$\ast$ and a transition 
\[
\begin{tikzcd}[column sep=1.5em]
a \quad : \quad
\ast
\ar[rr,"m"]
& {} &
\ast
\end{tikzcd}
\]
for each instruction $m$ of the machine displayed in
(\ref{equation/three-instructions})
parametrized by $x\in\Var$, $\valuev\in\Val$ and $\location\in\Loc$.
The graph~$H$ is moreover equipped with a "permutation tile"
\[
\,\,a\cdot b' \, \diamond_H \, b\cdot a'\,\,
\]
for every pair $a=a'$ and $b=b'$ of instructions of the machine.
The two asynchronous transition graphs~$(G,\diamond_G)$ and~$(H,\diamond_H)$ are
related by an asynchronous graph homomorphism
\begin{equation}\label{equation/asynchronous-graph-homomorphism-L}
\mathcal{L}  \quad : \quad (G,\diamond_G) \,\, \longrightarrow \,\, (H,\diamond_H)
\end{equation}
which maps every memory state $\memstate$ to the node~$\ast$, and every "instruction" to itself.
We recall the definition of such a homomorphism:
\begin{definition}[homomorphism]
An \emph{""asynchronous graph homomorphism""}
\begin{equation}\label{equation/asynchronous-graph-homomorphism}
\mathcal{F} \quad : \quad (G,\diamond_G) \,\, \longrightarrow \,\, (H,\diamond_H)
\end{equation}
is a graph homomorphism $\mathcal{F}:G\to H$ between the underlying graphs, such
that
\[
u\cdot v' \,\diamond_G\, v\cdot u'
\quad \Rightarrow \quad 
\mathcal{F}(u)\cdot \mathcal{F}(v')\,\diamond_H\,\mathcal{F}(v)\cdot \mathcal{F}(u')
\]
for all transitions $u$, $u'$, $v$, $v'$ of the asynchronous graph~$G$.
\end{definition}
\noindent
Note that, in that situation, one has
\[
f \sim g \quad \Rightarrow \quad \mathcal{L}(f)\approx\mathcal{L}(g)
\]
for all paths $f,g:M\transitionpath N$ in~$G$, where $\approx$ denotes the
permutation equivalence in the asynchronous graph~$(H,\diamond_H)$.

\paragraph{Data races as topological obstructions}
The reason for the liberal definition of $\diamond_H$ is that nothing should
forbid two instructions~$m_1$ and~$m_2$ to commute at the \emph{stateless} level
of abstraction.
By way of illustration, there exists a permutation tile in $H$ (depicted below
in light yellow) which permutes the two instructions $x:=2$ and $x:=3$ in the
following way:
\begin{equation}\label{equation/memoryless-tile-example}
\raisebox{-3em}{\includegraphics[width=8em]{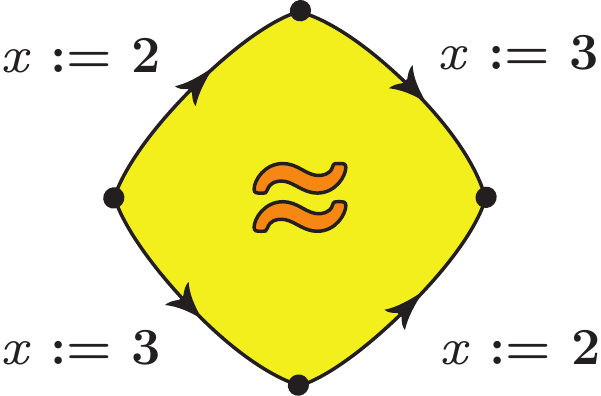}}
\end{equation}
%
This permutation tile~(\ref{equation/memoryless-tile-example}) should be
understood as a basic example of \emph{data race} in the machine, where the two
instructions $x:=2$ and $x:=3$ compete for the same variable~$x$.
As a matter of fact, one key observation and guiding idea of the paper is that
such a data race may be detected by the fact that it defines a permutation tile
in the stateless semantics~$(H,\diamond_H)$ which does \emph{not} lift
along~$\mathcal{L}$ to a permutation tile in the stateful
semantics~$(G,\diamond_G)$.
%
%
This line of thought leads us to the following definitions of 1-fibration and
2-fibration.
\begin{definition}[$1$-fibration]\label{def:1-fibration}
  An asynchronous graph homomorphism
  $\mathcal{F}:(G,\diamond_G)\to(H,\diamond_H)$ is called a \emph{""$1$-fibration""}
  when for every node~$x$ of~$G$ and transitions $v:\mathcal{F}(x)\to z$, there
  exists a transition $u:x\to y$ such that $\mathcal{F}(u)=v$.
\end{definition}
\begin{definition}[$2$-fibration]\label{def:2-fibration}
  An asynchronous graph homomorphism
  $\mathcal{F}:(G,\diamond_G)\to(H,\diamond_H)$ is called a \emph{""$2$-fibration""}
  when for every pair of transitions $u$ and $v'$ defining a path $u\cdot v'$ of
  length $2$ in~$G$ and for every "permutation tile"
\[
\mathcal{F}(u)\cdot\mathcal{F}(v') \,\diamond_H \, b \cdot a'
\]
  in~$H$, there exists a pair of transitions $v$ and $u'$ in~$G$ such that
\[
  u\cdot v' \diamond_G v \cdot u'
  \quad\mbox{and}\quad
  \mathcal{F}(v) = b
  \quad\mbox{and}\quad
  \mathcal{F}(u') = a'.
\]
%
\end{definition}
\noindent
Coming back to our construction, our point is that the asynchronous graph
homomorphism $\mathcal{L}$ defined
in~(\ref{equation/asynchronous-graph-homomorphism-L}) is \emph{not} a
"$2$-fibration" because of the presence of data races such
as~(\ref{equation/memoryless-tile-example}) in the stateless semantics.
Typically, any sequence of transitions in~$(G,\diamond_G)$
\begin{equation}\label{equation/x2x3inG}
\begin{tikzcd}[column sep=1.5em]
\memstate_1
\ar[rr,"{x:=2}"]
& {} &
\memstate_2
\ar[rr,"{x:=3}"]
&&
\memstate_3
\end{tikzcd}
\end{equation}
mapped by $\mathcal{L}$ to the upward border
$\begin{tikzcd}[column sep=1.4em]
\ast
\ar[rr,"{x:=2}"]
& {} &
\ast
\ar[rr,"{x:=3}"]
&&
\ast
\end{tikzcd}
$ of the permutation tile~(\ref{equation/memoryless-tile-example}) in the
asynchronous graph~$(H,\diamond_H)$ satisfies $\memstate_2(x)=2$ and
$\memstate_3(x)=3$.
For that reason, there exists no way to lift the permutation
tile~(\ref{equation/memoryless-tile-example}) along~$\mathcal{L}$ and to permute
the sequence of instructions~(\ref{equation/x2x3inG}) accordingly
in~$(G,\diamond_G)$ as follows:
\begin{equation}\label{equation/x3x2inG}
\begin{tikzcd}[column sep=1.5em]
\memstate_1
\ar[rr,"{x:=3}"]
& {} &
\memstate'_2
\ar[rr,"{x:=2}"]
&&
\memstate_3
\end{tikzcd}
\end{equation}
because this would mean that $\mu_3(x)=2$, and this would contradict the fact
that $\mu_3(x)=3$.
More generally, every data race in the machine may be detected as a topological
obstruction to the fact that the stateful-to-stateless
homomorphism~$\mathcal{L}$ is a "$2$-fibration".
Note that, in the same way but for different reasons, the data race between the
two instructions $x:=1$ described by the permutation tile in~$H$ below
\begin{equation}\label{equation/memoryless-tile-example-1-1}
\raisebox{-3em}{\includegraphics[width=8em]{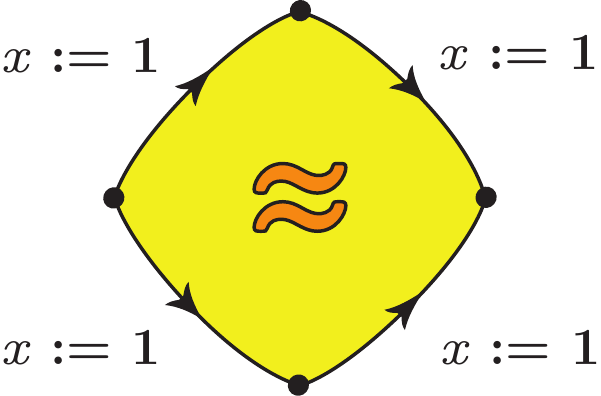}}
\end{equation}
does \emph{not} lift along $\mathcal{L}$ to a permutation tile in~$G$.
Indeed, the instruction $x:=1:\memstate\to\memstate'$ starting from any memory
state $\memstate$ has the nontrivial footprint $\writearea{x:=1}=\{x\}$, and is
thus not independent of itself in the asynchronous graph~$(G,\diamond_G)$.

\paragraph{An asynchronous semantics of code}
The machine just considered is a very elementary toy model, which can be easily
extended with locks and with memory allocation and deallocation.
Also, more than in the machine itself, we are interested in the asynchronous
description of the code~$C$ we want to analyse.
We thus need to explain how we shift from the machine to the code.
Interestingly, the story remains essentially the same.
To every program~$C$, we associate a stateful interpretation~$\semS{C}$ and a
stateless interpretation~$\semL{C}$ which reflect the interactive behavior of
the program~$C$ when confronted to its Environment, called Frame in that
context.
The two interpretations~$\semS{C}$ and~$\semL{C}$ are formulated as
\emph{"asynchronous transition systems"} ("ATS") related by a homomorphism
\begin{equation}\label{equation/lockmap-bis}
\begin{tikzcd}
   \mathcal{L}_C \quad : \quad \semS{C} \arrow[r] & \semL{C}
\end{tikzcd}
\end{equation}
mentioned in~(\ref{equation/lockmap}) which plays the same role for the code~$C$
as the homomorphism~(\ref{equation/asynchronous-graph-homomorphism-L}) for the
machine model.
The two "ATSs" $\semS{C}$ and $\semL{C}$ are defined uniformly by structural
induction on the program~$C$.
Their construction --- and more specifically the interpretation of the parallel
product $C_1\parallel C_2$ --- requires to develop a number of new techniques,
in particular an asynchronous "parallel product"
of two "ATSs" based on the same "machine model".

\paragraph{The asynchronous soundness theorem}
As in the case of the machine model, the data races produced by the program~$C$
will be detected as \emph{obstructions} to the fact that $\mathcal{L}_C$ is a
"$2$-fibration".
Typically, the program~$C$ defined as $ \,x:=2 \, ; \, x:=3\, $ is
data-race-free because the permutation tile
(\ref{equation/memoryless-tile-example}) does not appear in the stateless
semantics $\semL{C}$, while the program~$C'$ defined as $ \,x:=2 \parallel
x:=3\, $ produces a data race reflected by the fact that the permutation
tile~(\ref{equation/memoryless-tile-example}) appears in the stateless
interpretation~$\semL{C'}$ and cannot be lifted along~$\mathcal{L}$ to the
stateful interpretation~$\semS{C'}$.

\medbreak
In the present paper, we carry on our game-theoretic investigation of Concurrent
Separation Logic (CSL) initiated in~\cite{mfps} and establish that
\emph{well-specified} programs are data-race-free.
We achieve this by interpreting every derivation tree
\begin{equation}\label{equation/CSL-proof}
\prfsummary[\raiselabel{$\pi$}]{\prftree{\triple{\Gamma}{P}{C}{Q}}}
\end{equation}
of CSL as an asynchronous strategy $\semSep{\pi}$ playing on the asynchronous
game of separated states.
%
%
%
Our asynchronous version of the Soundness Theorem is then formulated in the
following fibrational way.
Suppose that a code~$C$ comes equipped with a "proof" of the "Hoare triple"
$\triple{\Gamma}{P}{C}{Q}$ in CSL, and consider the asynchronous subgraph
$\semSclosed{\{P\}C}{\tau}$ obtained by restricting $\semS{C}$ to the nodes
reachable from an "initial node" satisfying the precondition~$P$.
In that situation, we establish (see Thm.~\ref{thm/soundness} in
\S\ref{section/soundness} for details) that \medbreak
\noindent
\textbf{Asynchronous Soundness Theorem.} \emph{The stateful-to-stateless
  homomorphism $\mathcal{L}_{C}:\semS{C}\to\semL{C}$ is a
  "$2$-fibration" when restricted to the asynchronous
  subgraph~$\semSclosed{\{P\}C}{\tau}$.} \medbreak
\noindent
%
The "$2$-fibrational" property is conceptually new and provides the first structural
explanation for the absence of data races in concurrent programs specified
by~CSL.

\paragraph{Related works}
%
%
Stephen Brookes established the first proof of soundness of CSL
in~\cite{Brookes:a-semantics}, using a stateless trace semantics similar
to~$\semL{C}$ for the concurrent imperative programs.
%
More recently, Viktor Vafeiadis~\cite{Vafeiadis} gave a new proof of soundness,
based this time on a stateful operational semantics, similar to~$\semS{C}$.
Our approach can be seen as unifying the two schools of semantics, by revealing
the asynchronous graph morphism~(\ref{equation/lockmap-bis}) between them.
Also, one main benefit of our asynchronous approach is that we can directly
describe and analyze the concurrent execution of two instructions.

In the same way as we do here, Jonathan Hayman and Glynn
Winskel~\cite{hayman-Winskel} establish the soundness of CSL in a ``truly
concurrent'' setting. They interpret programs as Petri nets, where the
interference of the environment is modeled by adding events to the Petri net.
In contrast to our work, "precision" of the "invariants" is necessary for their
semantics to work whereas~\cite{Gotsman-precision} has shown that precision is
only needed in order to interpret properly the conjunction rule of CSL.
Another difference is that they consider a language somewhat different from
Brookes' original language~\cite{Brookes:a-semantics}, without local variables,
but with dynamic binding of resources.

We give in~\cite{mfps} a game-theoretic interpretation of CSL, where every Hoare
triple is interpreted as a \emph{game} between Adam and Eve, and every
derivation tree~$\pi$ as a \emph{winning strategy} for Eve in that game.
Every program is interpreted there as a set of purely sequential traces.
For that reason, it is not possible to establish in this framework the absence
of data races, at least in a nice and conceptual way.
One main achievement of the paper is thus to define a properly asynchronous game
semantics of CSL, and to derive for the first time the absence of data races
from purely semantic considerations on the model.
%

\paragraph{Synopsis of the paper}
After the machine states and instructions are described
in~\S\ref{sec:machine-states-machine-instructions}, we construct
in~\S\ref{sec:asynchronous-machine-models} the two asynchronous graphs
$\StateModelGraph$ and $\LockModelGraph$ defining our stateful and stateless
machine models.
We then explain in \S\ref{section/asc} how to interpret every code $C$ as a pair
$\semS{C}$ and $\semL{C}$ of "asynchronous transition systems" ("ATS") with
respective machine models $\StateModelGraph$ and $\LockModelGraph$.
Once the notions of "logical state" and of "separated state" are recalled
in~\S\ref{sec:logical-states} and in \S\ref{sec:separated-states}, we explain in
\S\ref{sec:semantics-of-proofs} how to interpret every proof $\pi$ of CSL as an
asynchronous strategy~$\semSep{\pi}$ playing on the machine
model~$\SepModelGraph$ of separated states.
From this, we establish our asynchronous soundness theorem in
\S\ref{section/soundness}, and conclude in \S\ref{section/conclusion}.

\section{Machine states and machine instructions}\label{sec:machine-states-machine-instructions}
We introduce below the notions of \emph{"machine state"} and of \emph{"machine
  instruction"} which will be used throughout the paper.
%
We suppose given countable sets $\Var$ of ""variable names"", $\Val$ of values,
$\Loc \subseteq \Val$ of ""memory locations"", and $\RVar$ of ""resources"".
In practice, we consider the case where $\Loc = \mathbb{N}$ and $\Val =
\mathbb{Z}$.

\begin{definition}[""Memory states""]
  A \emph{memory state} $\memstate$ is a pair $(s,h)$ of partial functions with
  finite domains $s: \Var \parfinmap \Val$ and $h: \Loc \parfinmap \Val$ called
  the \emph{stack}~$s$ and the \emph{heap}~$h$ of the memory state~$\memstate$.
  The set of memory states is denoted by $\State$.
  The domains of the partial function $s$ and of $h$ are denoted by
  $\vdom(\memstate)$ and $\hdom(\memstate)$ respectively, and we write
  $\dom(\memstate)$ for their disjoint union.
\end{definition}

\begin{definition}[""Machine states""]
  A \emph{machine state} is either a pair $\mstate=(\memstate,L)$ consisting of
  a "memory state" $\memstate$ and a subset of resources $L\subseteq\RVar$, called
  the \emph{lock state}, which describes the subset of locked resources in
  $\mstate$; or an error state $\Error$. The set of machine states is denoted by
  $\MStates$.
  Formally:
  \[
    \MStates = \State \times \powerset(\RVar) \,+\, \{\Error\}
  \]
\end{definition}
\noindent
A \emph{""machine step""} is defined as a labeled transition between "machine states".
There are two kinds of transitions:
\begin{equation}\label{equation/small-step-transition}
  \redto{(\memstate, L)}{m}{(\memstate', L')}
  \quad\quad\quad
  \redto{(\memstate, L)}{m}{\Error}
\end{equation}
depending on whether the instruction $m\in\Instr$ has been executed successfully
(on the left) or has produced a runtime error (on the right). In particular,
$\Error$ has no successor.
The \emph{""machine instructions""} $m\in\Instr$ which label the machine steps are
of the following form:
\begin{align*}
  m ::=\; &x \coloneqq E
      \mid x \coloneqq \deref{E}
      \mid \deref{E} \coloneqq E'
      \mid \knop\\
      \mid \;&x \coloneqq \kalloc(E,\location)
      \mid \kdispose(E)
      \mid P(r)
      \mid V(r)
\end{align*}
where $x\in\Var$ is a variable, $r\in\RVar$ is a "resource" name, $\location$~is a
"location", and $E,E'$ are arithmetic expressions, possibly with ``\emph{free}''
variables in~$\Var$.
For example, the instruction $x\coloneqq E$ executed in a "machine state"
$\mstate=(\memstate, L)$ assigns to the "variable"~$x$ the value $E(\mu)\in\Val$
when the value of the "expression" $E$ can be evaluated in the memory state~$\mu$,
and produces the runtime error~$\Error$ otherwise.
The instruction $P(r)$
acquires the "resource" variable $r$ when it is available, while the instruction
$V(r)$ releases it when~$r$ is locked, as described below:
\[
\begin{small}
\begin{array}{cc}
 \prftree{\evalexpr{E}{\memstate}{v}}
  {\redto{(\memstate, L)}{x \coloneqq E}{(\updmap{\memstate}{x}{v}, L)}}
&
 \prftree{E(\memstate)\,\, \text{not defined}}
  {\redto{(\memstate, L)}{x \coloneqq E}{\Error}}
  \vspace{.4em}
\\
    \prftree{r\notin{}L}
  {\redto{(\memstate, L)}{P(r)}{(\memstate, L\uplus\{r\})}}
&
    \prftree{r\notin{}L}
  {\redto{(\memstate, L\uplus\{r\})}{V(r)}{(\memstate, L)}}
  \end{array}
\end{small}
\]
The inclusion $\Loc \subseteq \Val$ means that an "expression" $E$ may also
denote a location.
In that case, $[E]$ refers to the value stored at location~$E$ in the heap.
The instruction $x \coloneqq \kalloc(E,\location)$ allocates some memory space
on the "heap@memory state" at address $\location\in\Loc$, initializes it with
the value of the expression $E$, and assigns the address $\location$ to the
variable~$x\in\Var$ if $location$ was free, otherwise there is no transition.
$\kdispose(E)$ deallocates the location denoted by~$E$ when it is allocated, and
returns~$\Error$ otherwise.
Finally, the instruction $\knop$ (for no-operation) does not alter the state.
%


\section{Asynchronous Machine Models}\label{sec:asynchronous-machine-models}
As explained in the introduction, \emph{machine models} are described using
asynchronous graphs.
Since we consider \emph{stateful} as well as \emph{stateless} descriptions of
the machine and of the code, we will consider two kinds of machine models,
organized into a pair of asynchronous graphs: the \emph{"stateful
  model"}~$\StateModelGraph$ based on "machine states", and the \emph{"stateless
  model"}~$\LockModelGraph$ based on "locks".
Their "tiles" will be defined using the notion of \emph{""footprint""}, which
summarizes which area of the "state@machine state" (memory, locks) an "instruction" relies on, and
how it uses it.
In both cases, we write $\footprint_s(m)$ for the footprint of an "instruction" $m$ in
state $s$, omitting the subscript when it is clear from the context.
Our machine models $\StateModelGraph$ and $\LockModelGraph$ are parameterized
over the finite set $\intro[well-defined locks]{\Locks}\subseteq\RVar$ of "locks", or "resources", which are
considered well-defined.
We sometimes write $\StateModelGraph(\Locks)$ or $\LockModelGraph(\Locks)$ to
make it explicit.
%
%
%

\paragraph{The stateful model}
A \emph{""machine state footprint""} 
\[
  \fp \in \powerset(\Var+\Loc) \times
  \powerset(\Var+\Loc) \times
  \powerset(\Locks) \times
  \powerset(\Loc)
\]
is, made of:
(i) $\readarea{\fp}$, the part of the memory that is \emph{read},
(ii) $\writearea{\fp}$, the part of the memory that is \emph{written},
(iii) $\lock(\fp)$, the locks that are \emph{touched},
and (iv) $\allocarea{\fp}$ the addresses that are \emph{allocated} or
\emph{deallocated}.
Two footprints $\fp$ and $\fp'$ are declared \emph{independent} when:
\[
\begin{array}{c}
(\,\readarea{\fp\phantom{'}} \cup \writearea{\fp\phantom{'}}\,) \cap \writearea{\fp'} \,\,=\, \emptyset
\\
(\,\readarea{\fp'} \cup \writearea{\fp'}\,) \cap \writearea{\fp\phantom{'}} \,\,=\, \emptyset
\end{array}
\quad\quad
\begin{array}{c}
\lock(\fp) \cap \lock(\fp') = \emptyset
\\
\allocarea{\fp} \cap \allocarea{\fp'} = \emptyset
\end{array}
\]
The \emph{""stateful model""} $\StateModelGraph$ is the following asynchronous
graph: its nodes are the machine states in $\MStates$, its transitions are of
the form
\[
  (\mu,L) \yrightarrow{m} (\mu',L')
  \quad \mbox{or} \quad
  (\mu,L) \yrightarrow{m} \Error
\]
corresponding to the machine steps, defined in
\S\ref{sec:machine-states-machine-instructions}.
The asynchronous tiles of $\StateModelGraph$ are the squares of the form
\[
  \mstate \yrightarrow{m} \mstate_1 \yrightarrow{m'} \mstate' \quad \sim
  \quad \mstate \yrightarrow{m'} \mstate_2 \yrightarrow{m} \mstate'
\]
where their footprints are independent in the sense above.
 
\paragraph{The stateless model}
A \emph{""lock footprint""}
\[
  \fp \in \powerset(\Locks) \times \powerset(\Loc)
\]
is made of a set of locks~$\lock(\fp)$ and a set of locations~$\allocarea{\fp}$.
%
Two such footprints are \emph{independent} when their sets are componentwise
disjoint.
The \emph{""stateless model""} $\LockModelGraph$ is defined in the following way:
its nodes are the subsets of $\Locks$, and its transitions are all the edges of
the form (note the non-determinism)
\[
\begin{array}{ccccc}
  L \xrightarrow{\,\,P(r)\,\,} L\!\uplus\!\{r\} 
    & \quad &
  L \xrightarrow{\,\,\kalloc(\location)\,\,} L 
  & \quad &
  L \xrightarrow{\;\;\tau\;\;} L 
\vspace{.2em}
\\
  L\!\uplus\!\{r\} \xrightarrow{\,\,V(r)\,\,} L
  &&
  L \xrightarrow{\,\,\kdispose(\location)\,\,} L 
  &&
   L \xrightarrow{\;\;\iwouldprefernotooverline{m}\;\;} \Error
\end{array}\]
where $\iwouldprefernotooverline{m}$ is a \emph{lock instruction} of the form:
\[
  P(r) \mid V(r) \mid \kalloc(\location) \mid \kdispose(\location) \mid \tau
\]
for $\location\in\Loc$ and $r\in\Locks$.
The purpose of these transitions is to extract from each instruction of the
machine its synchronization behavior.
An important special case, the transition~$\tau$ represents the absence of any
synchronization mechanism in an instruction like $x:=E$, $x:=[E]$ or $[E]:=E'$.
The asynchronous tiles of $\LockModelGraph$ are the squares of the form
\[
L \yrightarrow{x} L_1 \yrightarrow{y} L'
\quad 
\sim
\quad 
L \yrightarrow{y} L_2 \yrightarrow{x} L'
\]
when the lock footprints of $x$ and $y$ are independent.
It is worth noting that $L'$ may be equal to~$\Error$ in such an asynchronous
tile.
%
Note that the asynchronous graph $\LockModelGraph$ is more liberal than
$\StateModelGraph$ about which footprints commute, because it only takes into
account the locks as well as the allocated and deallocated locations.
As explained in the introduction, this mismatch enables us to detect \emph{data
  races} in the machine as well as in the code.

\paragraph{Remark}
The last component $\allocarea{\fp}$ in the machine state footprint as well as
in the lock footprint enables us to forbid a deallocation followed by an
allocation to happen at the same address without some kind of synchronization,
both at the stateful and stateless level.
This is consistent with practice, since the malloc implementation would
typically synchronize its accesses to the free-list(s) of the different threads.

\section{Asynchronous Semantics of Code}\label{section/asc}
In this section, we associate to every program~$C$ a pair of asynchronous
transition systems~$\semS{C}$ and $\semL{C}$ over the machine
models~$\StateModelGraph$ and~$\LockModelGraph$ introduced in the previous
section.
%
The first interpretation $\semS{C}$ is \emph{stateful} and describes how each
instruction of the program~$C$ acts on the memory states and on the locks.
The second interpretation $\semL{C}$ is \emph{stateless} and only remembers the
action of the instructions on the locks.
%
%
%
%
%

\subsection{Asynchronous transition systems (ATSs)}\label{section/ats}
"Asynchronous transition systems" ("ATSs") are specific asynchronous graphs where
every transition is either executed by Code or by Frame.
We thus start by introducing the following notion:
\begin{definition}[Asynchronous graph with polarities]
  An asynchronous graph with polarities is an asynchronous graph
  $(G,\diamond_G)$ where every transition is assigned a \emph{polarity} Code or
  Frame.
  One requires that in every permutation tile~${u\cdot v'\diamond_G
    v\cdot u'}$, the two transitions $u$ and $u'$ (symmetrically $v$ and $v'$) have the same polarity.
\end{definition}
A path in an asynchronous graph~$G$ with polarities is called \emph{Code-proper}
when it contains (at least) one Code transition.
A node~$x$ is called \emph{""initial""} in~$G$ when there are no Code-proper
incoming paths into~$x$, and \emph{""final""} when there are no Code-proper outgoing
paths from~$x$.
The sets of initial and final nodes in~$G$ are denoted~$\sourcefn{G}$
and~$\targetfn{G}$, respectively.
The graph~$G$ is called \emph{Code-acyclic} when there are no Code-proper
cycles, that is, every cycle of the graph~$G$ contains only Frame transitions.
A set $S$ of nodes of a graph is \emph{forward-closed} when $x\in S$ and $x\to
y$ implies that $y\in S$.

\begin{definition}[ATS]\label{def:ATS}
  An \emph{""asynchronous transition system""} (ATS) is a Code-acyclic asynchronous
  graph with polarities~$(G,\diamond_G)$ equipped with a forward-closed subset
  $\finalnodes{G}\subseteq\targetfn(G)$ of "final" nodes.
  A final node in $\finalnodes{G}$ is called a \emph{""returning node""} of the ATS.
\end{definition}
\begin{definition}\label{def:ATS-with-machine-model}
  An \emph{ATS} with machine model~$(\genFPgraph,\diamond)$ is defined as an
  ATS~$(G,|G|)$ equipped with an asynchronous graph homomorphism
  \[
    \getstate{G} \quad:\quad (\high{G},\diamond_G) \;\;\xrightarrow{\quad\quad}\;\; (\genFPgraph,\diamond)
  \]
%
One requires moreover that

\newcommand{\listeindent}{\hskip 6pt}

\noindent 
\listeindent \textbf{1}. the map $\getstate{G}$ defines a bijection between the
set~$\sourcefn{G}$ of "initial nodes" and the set of nodes of~$\genFPgraph$, and
an injection from the set~$\finalnodes{G}$ of "returning nodes" into the set of
nodes of~$\genFPgraph$.

\noindent 
\listeindent \textbf{2}. the map $\getstate{G}$ is a Frame "$1$-fibration", in the
sense that for every transition $v: \getstate{G}(x)\to z$ in the machine
model~$\genFPgraph$, there exists a unique Frame transition $u:x\to y$ in $G$
such that $\getstate{G}(u)=v:\getstate{G}(x)\to\getstate{G}(y)$,

\noindent
\listeindent \textbf{3}. the map $\getstate{G}$ is a Code-Frame and Frame-Frame
"$2$-fibration", in the sense that for every sequence of transitions
$\begin{small}
  \begin{tikzcd}[column sep=1.8em]
    x \ar[rr, "{u}"{description}] && y \ar[rr,"{v'}"{description}] && z
  \end{tikzcd}
\end{small}
$
in~$G$
%
%
where~$v':y\to z$ is a Frame transition, and for every "permutation tile"
in~$\genFPgraph$ of the form:
\begin{equation}\label{equation/two-fibration-tile-down}
\raisebox{-2.5em}{\includegraphics[height=5em]{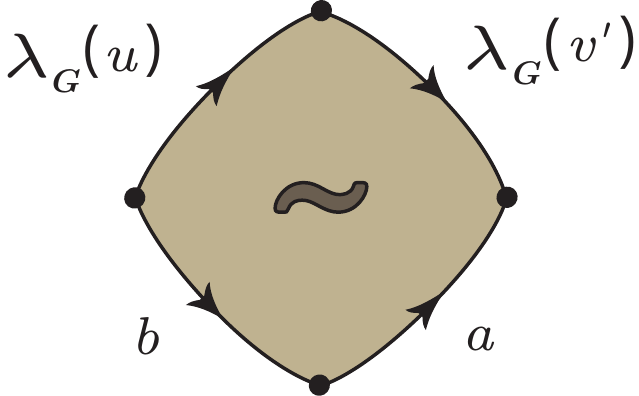}}
\end{equation}
there exists a sequence of transitions
$\begin{small}
  \begin{tikzcd}[column sep=1.8em]
x \ar[rr, "{v}"{description}] && y' \ar[rr,"{u'}"{description}] && z
\end{tikzcd}
\end{small}
$
and a "permutation tile" $u\cdot v'\diamond_G v\cdot u'$ in~$G$
transported by $\getstate{G}$ to the "permutation tile"
(\ref{equation/two-fibration-tile-down}) in the sense that
\[
\begin{tikzcd}[column sep=1.5em]
\getstate{G}(x) \ar[rr, "{\getstate{G}(v)}"] && \cdot \ar[rr,"{\getstate{G}(u')}"] && \getstate{G}(z) 
\end{tikzcd}
\,\, = \,\,
\begin{tikzcd}[column sep=1em]
\getstate{G}(x) \ar[rr, "b"] && \cdot  \ar[rr,"a"] && \getstate{G}(z)
\end{tikzcd}
\]
\end{definition}
%
%
\paragraph{Notation}
We often find convenient to label the transitions~$u:x\to y$ in~$G$ with the
instruction or lock instruction~$m$ which labels the
transition~$\getstate{G}(u)$ in the underlying asynchronous graph
$\genFPgraph_S$ or $\genFPgraph_L$.
We also write $\CODEbig{m}$ or $\ENVbig{m}$ to mean that the transition $u:x\to
y$ has the polarity Code or Frame in~$G$, respectively.

\subsection{Basic constructions on ATSs}\label{section/basic-operations}
The asynchronous interpretations $\semS{C}$ and $\semL{C}$ of the program~$C$
are performed by structural induction, using a number of primitive operations on
"ATSs" defined below.
Note that whenever a construction makes some nodes unreachable from the "initial
nodes", they are implicitly removed.
\paragraph{Sum}
The sum of two "ATSs" $G_1$ and $G_2$ with same machine model~$\genFPgraph$,
written $G_1 \juxt G_2$, is the disjoint union of the two asynchronous
graphs~$G_1$ and~$G_2$, where we identify their respective initial and returning
states together, when they have the same image under $\getstate{G_1}$ and
$\getstate{G_2}$.
This means that for the case of the "returning states", there are three cases.
If they both have returning states, we identify $\targetfn(G_1)$ with
$\targetfn(G_2)$;
if only one of $G_1$ and $G_2$ has returning states, we keep this one as our
returning states;
otherwise the juxtaposition has no returning states.

\paragraph{Sequential composition}
The \emph{""sequential composition""} $G \sequential G'$ of two "ATSs"~$G$ and $G'$ is
the disjoint union of~$G$ and $G'$ where we identify the "returning nodes" of~$G$
and the "initial nodes" of~$G'$ with the same underlying image under
$\getstate{G}$ and $\getstate{G'}$. Because we remove the inaccessible nodes,
when $G$ has no "returning nodes", $G\sequential G' = G$.
%

\paragraph{Parallel product}
The \emph{""parallel product""} $G_1\parprod~G_2$ of two "ATSs" $G_1$ and $G_2$ over
the same "machine model"~$\genFPgraph$ is defined as follows.
The nodes of $\high{G_1 \parprod G_2}$ are the pairs of nodes $x_1\,|\,x_2 \in
\high{G_1} \times \high{G_2}$ such that $\getstate{G_1}(x_1) =
\getstate{G_2}(x_2)$ and $\getstate{G_1\parprod G_2}(x_1,x_2)$ is defined to be
that common value.
%
The transitions of $\high{G_1 \parprod G_2}$ are of three kinds:
\medbreak

\noindent
1. the Code transitions 
$\,\,x_1| x_2 \yrightarrow{\CODE{m}} x_1'| x_2'\,\,$ where
\[
  x_1\yrightarrow{\CODE{m}} x_1' \text{ in $\high{G_1}$}
\quad\quad\text{and}\quad\quad
x_2 \yrightarrow{\ENV{m}} x_2'\text{ in $\high{G_2}$.}
\]

\noindent
2. the Code transitions 
$\,\,x_1| x_2 \yrightarrow{\CODE{m}} x_1'| x_2'\,\,$ where
\[
  x_1\yrightarrow{\ENV{m}} x_1'\text{ in $\high{G_1}$}
\quad\quad\text{and}\quad\quad
x_2 \yrightarrow{\CODE{m}} x_2' \text{ in $\high{G_2}$.}
\]

\noindent
3. the Frame transitions 
$\,\,x_1| x_2 \yrightarrow{\ENV{m}} x_1'| x_2'\,\,$ where
\[
  x_1\yrightarrow{\ENV{m}} x_1' \text{ in $\high{G_1}$}
\quad\quad\text{and}\quad\quad
x_2\yrightarrow{\ENV{m}} x_2' \text{ in $\high{G_2}$}.
\]
Note that every transition~$u:x_1|x_2\to y_1|y_2$ in the graph~$G_1\parallel
G_2$ is a pair $u=(u_1,u_2)$ also written $u=u_1|u_2$ of a
transition~$u_1:x_1\to y_1$ in $G_1$ and $u_2:x_2\to y_2$ in $G_2$.
%
%
%
A "permutation tile" in $G_1|G_2$

%
\begin{center}
\includegraphics[height=7em]{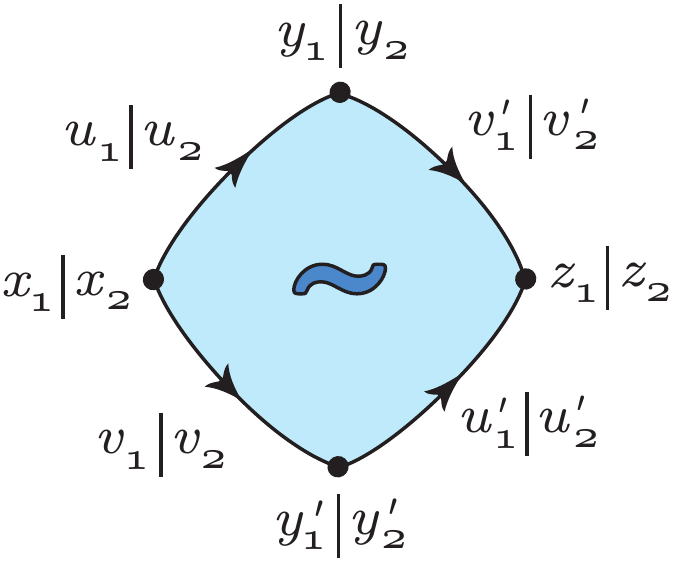}
\end{center}
is then defined as a square whose projections
\begin{center}
\includegraphics[height=7em]{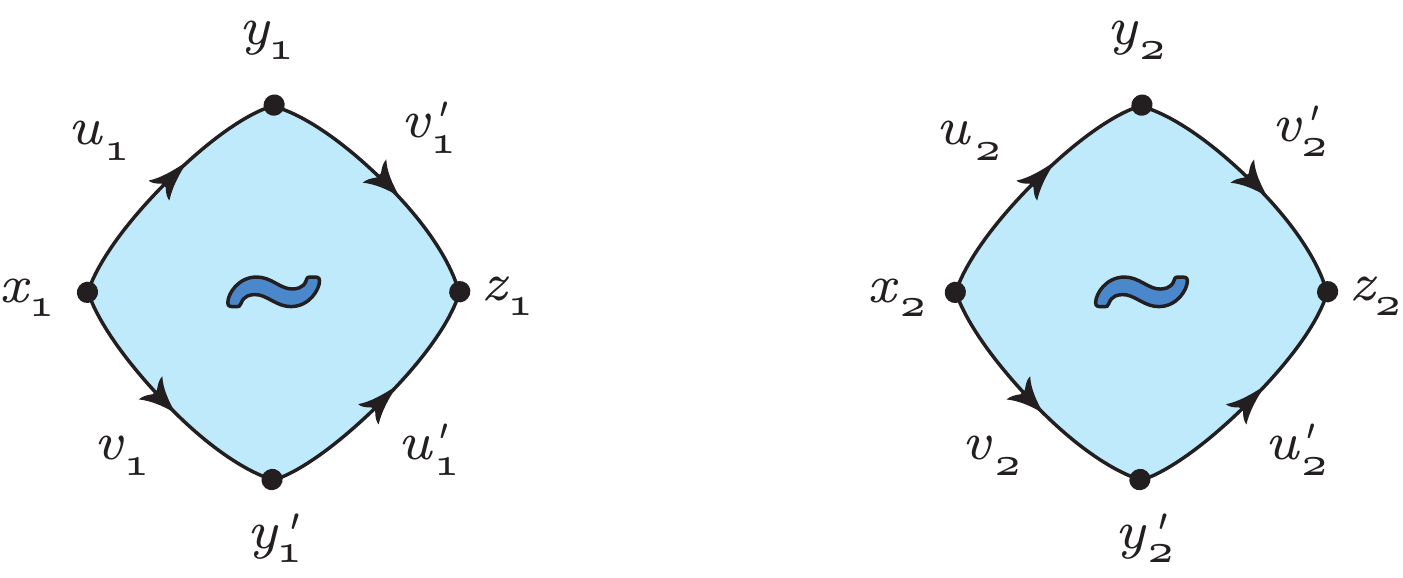}
\end{center}
%
\[
%
\]
define "permutation tiles" in $(G_1,\diamond_1)$ and $(G_2,\diamond_2)$,
respectively.
Finally, the "returning nodes" $x_1|x_2\in |G_1||G_2|$ are defined as the pairs
$x_1|x_2$ of "returning nodes" $x_1\in|G_1|$ and~$x_2\in|G_2|$.

The "parallel product" of~$G_1$ and~$G_2$ is \emph{asynchronous} in the sense that
every Code transition in $G_1\parallel G_2$ is a Code transition performed
by~$G_1$ and seen as a Frame transition by~$G_2$, or symmetrically, a Code
transition performed by~$G_2$ and seen as a Frame transition by~$G_1$.
In particular, by definition, the two components $G_1$ and $G_2$ never execute
(or ``fire'') a Code transition simultaneously in $G_1\parallel G_2$.
At the level of "permutation tiles", a Code transition $u_1|u_2:x_1|x_2\to
y_1|y_2$ performed in $G_1\parallel G_2$ by the component~$G_1$ and a Code
transition $v'_1|v'_2:y_1|y_2\to z_1|z_2$ performed in $G_1\parallel G_2$ by the
component~$G_2$ define a "permutation tile" precisely when the transitions
$\getstate{G_1\|G_2}(u_1|u_2)=\getstate{G_1}(u_1)=\getstate{G_2}(u_2)$ and
$\getstate{G_1\|G_2}(v'_1|v'_2)=\getstate{G_1}(v'_1)=\getstate{G_2}(v'_2)$
define a "permutation tile" in the underlying machine model~$\genFPgraph$.
As a matter of fact, one purpose of the "machine model"~$(\genFPgraph,\diamond)$
is precisely to provide that piece of information necessary to construct the
"parallel product" of~$G_1$ and~$G_2$.

%
%

%
%
%
%



\paragraph{Resource hiding}
In order to interpret the "resource" introduction construct $\resource{r}{C}$, we
introduce a \emph{hiding} operator $\hide{r}$ on "ATSs" which hides the new
resource $r$, similarly to the operator $\nu$ in the $\pi$-calculus.
Formally, if $G$ is an "ATS" over $\genFPgraph(\Locks\uplus\!\{r\})$, then
$\hide{r}(G)$ is the "ATS" over $\genFPgraph(\Locks)$ where: (1) the "resource"~$r$
has been removed from the sets of locked resources of all states, (2) the Code
transitions $P(r)$ and $V(r)$ are replaced with $\knop$s, (3) the Frame
transitions $P(r)$ and $V(r)$ are removed from the graph~$G$, and (4) the
remaining "permutation tiles" are preserved.
Moreover, we only keep as initial and "returning states" the initial and returning
states~$x$ of $G$ such that the resource~$r$ is not held in $\getstate{G}(x)$.
%
%
%

\paragraph{Critical sections}
Dually, inside critical sections, we need to ``lift'' ATSs over some set
$\Locks$ of locks to ATSs over $\Locks\uplus\{r\}$.
This can be done naturally in this case because we know that, during the critical
section, the resource $r$ is held by the Code.
Formally, $\ATSwhen{r}(G)$ has the same underlying asynchronous graph as $G$,
where $\getstate{}' := \getstate{\ATSwhen{r}(G)}$ is defined by:
\begin{align*}
  \quad\quad\getstate{}'(x) \quad&:=\quad L \uplus \{r\} &&\text{ if $\getstate{G}(x) = L$}\\
  \quad\quad\getstate{}'(x) \quad&:=\quad (\memstate, L \uplus \{r\}) &&\text{ if $\getstate{G}(x) = (\memstate, L)$}.
\end{align*}
This does not define an ATS yet, for condition~$\mathbf{3}$ is not satisfied:
there are not enough Environment transitions. This is why we must freely add
Frame transitions and new nodes to make it an ATS. The returning nodes are
defined to be the same~as~$G$.

\paragraph{Other constructions on ATSs}
Given an ATS $G$ and a Boolean formula $B$, we define $\whentrue{B}(G)$ as the
graph $G$ where, among the Code transitions out of "initial nodes", we only keep
those where $B$ holds on $\getstate{G}(x)$. Then, we remove the nodes made
unreachable by this edge removal.
Similarly, we define $\whenfalse{B}$ for when $B$ does not hold.
Finally, $\whenabort{B}$ is the graph with transitions from the "initial states"
where $B$ errors out, because it tries to read undefined "variables",
to~$\Error$.
Note that in the case of the $\semL{C}$ semantics, since the nodes do not
contain information on the state, the first two constructions above are the
identity. This means that we sometimes consider impossible branches.

\subsection{Asynchronous semantics of the code}
We explain how to give a semantics to any code $C$ as an ATS $\sem{C}$, by
induction on its structure. This lets us build $\semS{C}$ and $\semL{C}$ in the
same way.
First, we give the syntax of our imperative concurrent language, which we
borrow from~\cite{Brookes:a-semantics,Vafeiadis}.
\begin{align*}
  B &::=   \True
      \mid \False
      \mid B \wedge B'
      \mid B \vee B'
      \mid E = E' \\
  E &::=   0
      \mid 1
      \mid \ldots
      \mid x
      \mid E + E'
      \mid E * E'\\
  C &::=   x \coloneqq E
      \mid x \coloneqq \deref{E}
      \mid \deref{E} \coloneqq E'
      \mid C ; C'
      \mid C_1 \parallel C_2
      \mid \kskip\\
    &\mid \while{B}{C}
      \mid \resource{r}{C}
      \mid \when{r}{B}{C}\\
    &\mid \ifte{B}{C_1}{C_2}
      \mid x \coloneqq \kmalloc(E)
      \mid \kdispose(E)
\end{align*}

\paragraph{Semantics of instructions}
To every "instruction"~$m \in \Instr$, we associate the "ATS"~$\sem{m}$ with "machine
model"~$\genFPgraph$ defined as two copies $\genFPgraph_0+\genFPgraph_1$ (called
\emph{source} and \emph{target}) of the "asynchronous graph"~$\genFPgraph$.
Every transition in $\genFPgraph_0+\genFPgraph_1$ is assigned the Frame
polarity.
To this, one adds a Code transition $x_0\to y_1$ for every transition of the
form~(\ref{equation/small-step-transition}) labeled by~$m$ in "the small step
semantics@machine step".
Here, $x_0$ and $y_1$ are the nodes $x$ and $y$ of~$\genFPgraph$ taken in the
source and target components $\genFPgraph_0$ and $\genFPgraph_1$ of~$\sem{m}$,
respectively.
The transition $x_0\to y_1$ is mapped by $\getstate{\sem{m}}$ to the transition
associated to the small step transition~(\ref{equation/small-step-transition})
in~$\genFPgraph=\StateModelGraph$ or~$\genFPgraph=\LockModelGraph$.
Finally, one adds a Code-Frame "permutation tile" in $\sem{m}$ for each Code-Frame
"permutation tile" in~$\genFPgraph$, in such a way that
$\getstate{\sem{m}}:\sem{m}\to\genFPgraph$ defines a Code-Frame "$2$-fibration".

\paragraph{Leaf codes}
For leaf codes that correspond to instructions (all, except for $\kmalloc$),
their semantics is the same as that of the instruction.
For $\kmalloc(E)$, we take the non-deterministic union of all the
$\kalloc(E,\location)$:
\[
  \sem{\kmalloc(E)} \quad := \quad \bigoplus_{\ell\in\Loc}\: \sem{\kalloc(E,\location)}
\]

\paragraph{Conditionals} Conditional branching is interpreted as
\begin{align*}
  \sem{\ifte{B}{C}{C_1}} &= \whentrue{B}(\sem{\knop})\,;\sem{C_1}\\
                            &\;\juxt \whenfalse{B}(\sem{\knop})\,;\sem{C_2}\\
                            &\;\juxt \whenabort{B}
\end{align*}
The $\knop$s are needed because the environment can interfere between the
evaluation of $B$ and the beginning of the execution of~$C_i$.

\paragraph{Sequential and parallel compositions}
We use the "sequential@sequential product" and "parallel product" of ATSs with machine models
defined in~\S\ref{section/basic-operations}, in the following way:
\[
  \sem{C_1\|C_2} \hspace{.5em} = \hspace{.5em} \sem{C_1}\|\sem{C_2}
  \quad\quad
   \sem{C_1;C_2} \hspace{.5em} = \hspace{.5em} \sem{C_1};\sem{C_2}.
\]

\paragraph{Resource introduction}
The interpretation of $\resource{r}{C}$ is defined as
\[
  \sem{\resource{r}{C}} \quad=\quad \hide{r}\big(\sem{C}\big)
\]

\paragraph{Critical sections}
The semantics $\sem{\when{r}{B}{C}}$ is defined using the "sequential composition"
above and $\mathrm{whentrue}$:
\[
  \whentrue{B}\Big(\sem{P(r)} ; \ATSwhen{r}\big(\sem{C}\big) ;\sem{V(r)}\Big) \juxt \whenabort{B}.
\]

\paragraph{Loops}
For loops, the interpretation of $C' = \while{B}{C}$ is defined as the (possibly
infinite) least fixpoint of the function $F$:
\begin{align*}
  F(G) &= \whentrue{B}\big(\sem{\knop}\big) \sequential \sem{C}\sequential G
           \juxt \whenfalse{B}\big(\sem{\knop}\big) \\
         &\quad\juxt \whenabort{B}.
\end{align*}
%

\paragraph{Remark}
The map $\getstate{G}$ is a "$2$-fibration" for Code-Frame and Frame-Frame
permutations, but not for Code-Code permutations in general.
Consider for instance the interpretation of the program
\[
C = \resource{r}{\big\{\,(P(r);V(r))\,\, \| \,\, (P(r);V(r)) \,\big\}}
\]
where $r\in\RVar$ is a resource name.
Since the resource introduction performed by $\resource{r}{C}$ is interpreted by
hiding the resource $r$, the two instructions~$P(r)$ and~$V(r)$ are both
transformed in $\knop$s instructions.
However the two $\knop$\emph{s do not} form a "tile"!
Another example is, of course, the "sequential composition"~$C_1;C_2$ of two
codes~$C_1$ and $C_2$.

\subsection{Comparing the stateful and the stateless semantics}\label{sec:comparing-semantics}
We construct a category of "ATSs" with machine models, in the following way.
A \emph{morphism} between "ATSs" with machine models
\[
\getstate{G_1}: \high{G_1} \longrightarrow \genFPgraph_1
\quad\quad\quad \getstate{G_2}: \high{G_2} \longrightarrow \genFPgraph_2
\]
is a
pair of "asynchronous graph morphisms" $\mathcal{F}:\genFPgraph_1\to\genFPgraph_2$
and $\mathcal{G}:G_1\to G_2$ such that the diagram below commutes:
\[
  \begin{tikzcd}[column sep = 6em, row sep=1.8em]
    \high{G_1} \arrow[r,"\mathcal{G}"{description}] \arrow[d,"{\getstate{G_1}\,\,}"{swap}] & \high{G_2}
    \arrow[d,"{\,\,\getstate{G_2}}"]\\
    \genFPgraph_1 \arrow[r,"\mathcal{F}"{description}] & \genFPgraph_2
  \end{tikzcd}
\]
One requires moreover that~$\high{\mathcal{G}}$ send "initial" (resp. "returning@returning node") nodes of $G_1$ 
to "initial" (resp. "returning@returning node") nodes of $G_2$.
This defines a category noted $\ATS$.
%
%
Let $\mathcal{F}:\StateModelGraph\to\LockModelGraph$ denote the asynchronous
graph morphism which transports every machine state $\mstate=(\memstate,L)$ to
the underlying subset $L\subseteq\Locks$ of locks held in $\mstate$.
%
Every "instruction" $m\in\Instr$ comes equipped with an ATS morphism
\[
  \semmorph_m=(\mathcal{F},\mathcal{G}_m) \quad : \quad \semS{m} \; \xrightarrow{\quad\quad} \semL{m}
\]
where the asynchronous graph morphism $\mathcal{G}_m$ is defined as
\begin{gather*}
  (\memstate,L) \xrightarrow{m} (\memstate',L') \quad \xmapsto{\quad\quad}%
  \quad L \xrightarrow{m} L'
\end{gather*}
%
%
Because the stateful and stateless interpretations~$\semS{-}$ and~$\semL{-}$ are
defined using the same functorial operations over~$\mathcal{F}$, we can
associate to every code~$C$ a morphism of $\ATS$
\[
  \semmorph_C=(\mathcal{F},\mathcal{G}_C) \quad : \quad \semS{C} \; \xrightarrow{\quad\quad} \semL{C}
\]
starting from the family of morphisms $\semmorph_m$ associated to instructions.
Note that this morphism $\semmorph_C=(\mathcal{F},\mathcal{G}_C)$ living in the
category~$\ATS$ plays a fundamental role in the present work, since our
asynchronous refinement of the original Soundness Theorem for CSL relies on it,
see \S\ref{section/soundness} for details.


\section{Logical States}\label{sec:logical-states}
As discussed in~\cite{mfps}, reasoning about concurrent programs in separation
logic requires to introduce an appropriate notion of \emph{"logical state"},
including information about "permissions".
The version of concurrent separation logic we consider is almost the same as its
original formulation by O'Hearn and Brookes~\cite{OHearn,Brookes:a-semantics}.
One difference is that we benefit from the work of Bornat, Calcagno, O'Hearn,
Parkinson and Yang in~\cite{Bornat:permissions,Bornat:sep,Bornat:hoare} and use
permissions~$p$ and the predicate $\own_p(x)$ in order to handle the heap as
well as variables in the stack.
We suppose given an arbitrary partial cancellative commutative monoid $\Perm$
which we call the \emph{""permission monoid""}, following \cite{Bornat:permissions}.
The element $\top$ will be used as the permission required for a program to
write somewhere in memory.
We thus require that~$\top$ does not admit any multiples, ie. $ \forall
x\in\Perm, \top \cdot x$ is not defined.
The intuition (which we will need to turn into a theorem) is that we prevent in
this way concurrent mutation and observation of the same "location", that is, data
races.
The set $\LStates$ of ""logical states"" is defined in much the same way as the set
$\State$ of "memory states", with the addition of "permissions":
\[
  \LStates \;=\; (\Var \parfinmap (\Val \times \Perm)) \times
  (\Loc \parfinmap (\Val \times \Perm))
\]
%
The main benefit of permissions is that they enable us to define a
\emph{separation product} $\state * \state'$ between two logical states $\state$
and $\state'$, which generalizes the disjoint union.
When it is defined, the logical state $\state * \state'$ is defined as a partial
function with domain
\[
\dom(\state * \state) = \dom(\state) \cup \dom(\state')
\]
in the following way: for $a\in\Var\amalg\Loc$,
\[
  \state*\state'(a) =
  \begin{cases*}
   \,\, \state(a) & if $a \in \dom(\state)\setminus\dom(\state')$\\
    \,\, \state'(a) & if $a \in \dom(\state')\setminus\dom(\state)$\\
   \,\, (v,p\cdot{}p') & if $\state(a) = (v, p)$ and $\state'(a) = (v,p')$\\
  \end{cases*}
\]
The separation product $\state * \state'$ of the two "logical states" $\state$ and
$\state'$ is not defined otherwise.
In particular, the "memory states" underlying $\state$ and $\state'$ agree on the
values of the shared variables and heap locations when the separation product is
well defined.
The syntax and the semantics of the ""formulas"" of Concurrent Separation Logic is
the same as in Separation Logic.
%
%
The grammar of formulas is:
\begin{align*}
  P,Q,R,J \Coloneqq \; &\emp \mid \True \mid \False \mid P \vee Q \mid
                         P \wedge Q \mid \neg P
  \mid\; \forall v. P
         \mid \; \exists v. P\\
  \mid\; &P * Q \mid v \pointsto{p} w \mid \own_p(x) \mid E_1=E_2
\end{align*}
where $x\in\Var$, $p\in\Perm$, $v,w\in\Val$.
Given a "logical state" $\sigma=(s,h)$ consisting of a logical stack~$s$ and of a
logical heap~$h$, the semantics of the "formulas", expressed as the predicate
$\state \vDash P$, is standard:
\begin{align*}
  \s \vDash v \pointsto p w \;&\Longleftrightarrow\; v\in\Loc \wedge s =\emptyset \wedge h = [v \mapsto (w, p)] \\
  \s \vDash \own_p(x) \;&\Longleftrightarrow\; \exists v\in\Val,  s = [x \mapsto (v, p)] \wedge h = \emptyset\\
  \s \vDash E_1 = E_2 \;&\Longleftrightarrow\; \sem{E_1} = \sem{E_2} \wedge \fv(E_1=E_2) \subseteq \vdom(s)\\
  \s \vDash P \wedge Q \;&\Longleftrightarrow\; \s \vDash P \text{ and } \s \vDash Q\\
  \s \vDash P*Q \;&\Longleftrightarrow\; \exists \s_1 \s_2, \, \s = \s_1 * \s_2 \text{ and } \s_1 \vDash P \text{ and } \s_2 \vDash Q.
\end{align*}
%
The ""proof system@proof"" underlying concurrent separation logic is a sequent calculus,
whose sequents are ""Hoare triples"" of the form
\[
\,\,\triplestd\,\,
\]
where $C\in\Code$, $P$, $Q$ are "predicates", and $\Gamma$ is a context, defined
as a partial function with finite domain from the set~$\RVar$ of "resource
variables" to predicates.
Intuitively, the context $\Gamma=r_1:J_1,\dots,r_k:J_k$ describes the ""invariant""
$J_i$ satisfied by the "resource variable" $r_i$.
The purpose of these "resources" is to describe the fragments of memory shared
between the various threads during the execution.

The inference rules of CSL are given in Figure~\ref{fig:inference-rules}.
The inference rule $\RES$ associated to $\resource{r}{C}$ moves a piece of
"logical state" which is owned by the Code into the shared context~$\Gamma$, which
means that it can be accessed concurrently inside the code~$C$.
However, the access to that piece of state is mediated by the $\kwith$
construct, which grants temporary access under the condition that one must give
it back (rule $\WHEN$).
Note that the rule \WHEN{} has the side condition $P \Rightarrow
\defpred(B)$. 
This means that if $P$ is true in some "logical state", then it implies, for each
free variable $x$ of $B$, that there exists some "permission" $p$ such that
$\own_p(x)$ holds.

Notice that the context $\Gamma=r_1:J_1,\dots,r_k:J_k$ is
required to be \emph{"precise"} in the rule \CONJ.
This means that each of the predicates $J_i$ is "precise" in the following sense:
\begin{definition}[Precise predicate]
  A predicate $P$ is \emph{""precise""} when, for every logical state
  $\state\in\LStates$, there exists at most one logical state $\state'\in\LStates$ 
  such that $\state' \vDash P$ and
\[
\exists \state''\in\LStates,\quad \state=\state'*\state''.
\]
\end{definition}

\begin{figure*} 
  \centering
  \small
\begin{mathpar}
  \prfbyaxiom{\AFF}{\triple{\Gamma}{(\own_\top(x) * P) \wedge E = v}{x \coloneqq
      E}{(\own_\top(x) * P) \wedge x = v}}
  \and
\prfbyaxiom{\STORE}
{\triple
  {\Gamma}
  {E \mapsto -}
  {\deref{E} \coloneqq E'}
  {E \mapsto E'}}
  \and
\prftree[r]{\LOAD}
{x \notin \fv(E)}
{\triple
  {\Gamma}
  {E \mapsto^{p} v * \own_{\top}(x)} 
  {x \coloneqq \deref{E}}
  {E \mapsto^{p} v * \own_{\top}(x) * x = v}}
  \and
  \prftree[r]{\SEQ}{\triple{\Gamma}{P}{C_1}{Q}}{\triple{\Gamma}{Q}{C_2}{R}}{\triple{\Gamma}{P}{C_1;C_2}{R}}
\and

\prftree[r]{\IF}
     {P \Rightarrow \defpred(B)}
     {\triple{\Gamma}{P\wedge B}{C_1}{Q}} 
     {\triple{\Gamma}{P\wedge \neg B}{C_2}{Q}} 
     {\triple{\Gamma}{P}{\ifte{B}{C_1}{C_2}}{Q}}
     \and
     \prftree[r]{\CONJ}{\Gamma\text{ is "precise"}}{\triple{\Gamma}{P_1}{C}{Q_1}}{\triple{\Gamma}{P_2}{C}{Q_2}}{\triple{\Gamma}{P_1
      \wedge P_2}{C}{Q_1 \wedge Q_2}}
  \and
  \prftree[r]{\DISJ}{\triple{\Gamma}{P_1}{C}{Q_1}}{\triple{\Gamma}{P_2}{C}{Q_2}}{\triple{\Gamma}{P_1
      \vee P_2}{C}{Q_1 \vee Q_2}}
  \and
  \prftree[r]{\RES}{\triple{\Gamma,
      r:J}{P}{C}{Q}}{\triple{\Gamma}{P*J}{\resource{r}{C}}{Q*J}}
  \and
  \prftree[r]{\WHEN}{P\Rightarrow \defpred(B)}{\triple{\Gamma}{(P*J)\wedge
      B}{C}{Q*J}}{\triple{\Gamma, r:J}{P}{\when{r}{B}{C}}{Q}}
  \and
  \prftree[r]{\PAR}{\triple{\Gamma}{P_1}{C_1}{Q_1}}{\triple{\Gamma}{P_2}{C_2}{Q_2}}{\triple{\Gamma}{P_1*P_2}{C_1\parallel
      C_2}{Q_1 * Q_2}}
  \and
  \prftree[r]{\FRAME}
  {\triple{\Gamma}{P}{C}{Q}}
  {\triple{\Gamma}{P*R}{C}{Q*R}}
\end{mathpar}
\normalsize
\caption{Inference rules of Concurrent Separation Logic}
\label{fig:inference-rules}
\end{figure*}

\section{The machine model of separated states}\label{sec:separated-states}
We recall the notion of \emph{"separated state"} formulated in~\cite{mfps} whose
purpose is to separate the logical memory state into one region controlled by
the Code, one region controlled by the Frame, and one independent region for
each unlocked "resource".
In order to define the notion, we suppose given a finite set~$\Locks \subseteq
\RVar$ of resource variables, or locks.
\begin{definition}\label{def:separated-states}
  A \emph{""separated state""} is a triple
    \[
    (\state_C, \statevector, \state_F) \in \LStates \times (\Locks \to \LStates+\{C,F\}) \times \LStates
  \]
such that the logical state below is defined:
\begin{equation}\label{eqn:sep-state-product}
  \state_C \, * \,\, \Big\{\,\,\bigsprod_{r \in \dom(\statevector)}\statevector(r) \,\, \Big\}\,\, * \, \state_F
  \quad \in \quad \LStates
\end{equation}
where 
\begin{align*}
    \dom(\statevector)  &= \{\,r\in\Locks \mid
                              \statevector(r)\in\LStates\hspace{0.1em}\},
    \\
    \domC(\statevector) &= \{\,r\in\Locks \mid \statevector(r)=C\hspace{0.1em}\},
    \\
    \domF(\statevector) &= \{\,r\in\Locks \mid \statevector(r)=F\hspace{0.1em}\}.
\end{align*}
\end{definition}
\noindent
We say that a separated state $(\state_C, \statevector, \state_F)$ combines into
a machine state $\mstate = (\memstate, L)$ precisely when $L =
\domC(\statevector)\uplus\domF(\statevector)$ and when the function
$U:\LStates\to\State$ which forgets the permissions transports the logical
state~(\ref{eqn:sep-state-product}) into the memory state~$\memstate\in\State$.
Note that, by definition, every separated state $(\state_C, \statevector,
\state_F)$ combines into a unique machine state, which we write for concision
\begin{equation}\label{equation/linking}
(\memstate,L) \quad = \quad \mbox{$\bigsprod$}(\state_C, \statevector, \state_F).
\end{equation}
Interestingly, the notion of separated state comes with the same notion of
"footprint@machine state footprint" as the machine states, defined as elements of
\[
  \fp \in \powerset(\Var+\Loc) \times \powerset(\Var+\Loc) \times
  \powerset(\Locks) \times \powerset(\Loc).
\]
which describes the footprint of a transition by Eve or Adam.
\begin{definition}\label{def:separation-graph}
  The \emph{""machine model of separated states""} $\SepModelGraph$ is the
  asynchronous graph whose nodes are the separated states and whose edges are
  either Adam or Eve transitions:

  \smallskip
  \noindent
  $\;\bullet\;$ Eve transitions are of the form
    \[
      (\state_C, \statevector, \state_F) \yrightarrow{\,\,\CODE{m}\,\,} (\state'_C, \statevector', \state_F)
    \]
    where~$m\in\Instr$ is an instruction such that
\[
\redto{\mbox{$\bigsprod$}(\state_C, \statevector, \state_F)}{m}{\mbox{$\bigsprod$}(\state'_C, \statevector', \state_F)}
\]
    and such that the following conditions are
    satisfied:
    \begin{align*}
     \forall \location\notin&\writearea{m},\;\state_C(\location) = \state'_C(\location) &
      \writearea{m} \cup \readarea{m}&\subseteq \dom(\state_C)\\
      \lock(m) &\subseteq \dom(\statevector) \cup \domC(\statevector) &
      \forall r\notin\lock(m),\; &\statevector(r) = \statevector'(r).
    \end{align*}
  $\;\bullet\;$ Adam moves of the form
    \[
      (\state_C, \statevector, \state_F) \xrightarrow{\,\,\ENV{m}\,\,}
      (\state_C, \statevector', \state'_F)
    \]
    where~$m\in\Instr$ is an instruction, such that 
\[
\redto{\mbox{$\bigsprod$}(\state_C, \statevector, \state_F)}{m}{\mbox{$\bigsprod$}(\state_C, \statevector', \state'_F)}
\]
and such that the following conditions are
satisfied: 
    \begin{align*}
     \forall \location\notin&\writearea{m},\;\state_F(\location) = \state'_F(\location) &
      \writearea{m} \cup \readarea{m}&\subseteq \dom(\state_F)\\
      \lock(m) &\subseteq \dom(\statevector) \cup \domF(\statevector) &
      \forall r\notin\lock(m),\; &\statevector(r) = \statevector'(r).
    \end{align*}
\end{definition}

\medskip

\noindent
Like the two other machine models~$\StateModelGraph$ and~$\LockModelGraph$,
the "tiles" of $\SepModelGraph$ are the squares 
\begin{center}
 \squarediag{x}%
 {z}%
 {y}%
 {y'}%
 {m}{m'}{m'}{m}
\end{center}
where the footprints of $m$ and $m'$ at state $x$ are independent.
More concretely, an Eve-Eve "tile" is of the following form (where we only write
the first component of each separated state): 
\begin{center}
 \squarediag{\state_1 * \state_2 * \state}%
 {\state'_1 * \state'_2 * \state}%
 {\state'_1 * \state_2 * \state }%
 {\state_1 * \state'_2 * \state}%
 {m}{m'}{m'}{m}
\end{center}
For example, the first state $\state_C$ is split into
$\state_1*\state_2*\state$, where the domain of $\state_1$ is $\writearea{m}$,
that of $\state_2$ is $\writearea{m'}$ and $\state$ is the rest of $\state_C$.
The resulting definition of the machine model $\SepModelGraph$ of separated
states ensures that the operation~(\ref{equation/linking}) defines a morphism
$\circledast:\SepModelGraph\to\StateModelGraph$ of asynchronous graphs from
$\SepModelGraph$ to the stateful model~$\StateModelGraph$.

%

%
%


\section{An asynchronous semantics of proofs}\label{sec:semantics-of-proofs}
In this section, we interpret derivation trees (or proofs) of CSL in our
asynchronous semantics.
In the same way as we did for the Code in \S\ref{section/asc}, we interpret
every proof $\pi$ of a "Hoare triple" $\triplestd$ as an "asynchronous transition
system" ("ATS", Definition~\ref{def:ATS}).
The underlying asynchronous machine model is $\SepModelGraph$, 
the graph of separated states.
As in the previous case, our ATSs have the "$1$-fibration" property, and moreover
the initial states are all the states that satisfy~$P$, and all the "final states"
satisfy~$Q$.
The interpretation $\semSep{\pi}$ also satisfies that the second component
$\statevector$ of all its nodes of $\semSep{\pi}$ satisfies the "invariants" of
$\Gamma$ pointwise.
In order to define the interpretation of a proof $\pi$ by induction on its
structure, we start by defining a small number of new constructions on ATSs.

\paragraph{The parallel product with separated states}
In order to define the "parallel product" $G_1\|G_2$ of two ATSs on the model
$\SepModelGraph$ of separated states, we need to adapt the compatibility
condition given by the equality $\getstate{G}(x_1) = \getstate{G}(x_2)$ in the
case of the stateful and stateless models~$\StateModelGraph$
and~$\LockModelGraph$.
In the case of $\SepModelGraph$, two nodes of $G_1$ and $G_2$ should be declared
compatible when they describe the two ``subjective'' (and dual) views of the
same situation, provided in this case by a separated state of $G_1\|G_2$.
%
This leads us to the notion of \emph{three-party} separated state, defined as a
tuple $(\state_1, \state_2, \statevectorstar, \state_F)$ where $\state_1,
\state_2, \state_F \in \LStates$ are logical states, where $\statevectorstar :
\Locks \to \LStates+\{C_1,C_2,F\}$, and where the product $\circledast
(\state_1, \state_2, \statevectorstar, \state_F)$ immediately adapted
from~(\ref{eqn:sep-state-product}) is well-defined.

We define three functions on these new separated states:
the ``objective'' projection, which corresponds to the view of the whole program
$C_1\parallel C_2$, is defined by:
\[
  (\state_1, \state_2, \statevectorstar, \state_C)
  \mapsto
  (\state_1 * \state_2, \statevectorstar[C_i\mapsto C], \state_C)
\]
the left and right ``subjective'' projections
\begin{align*}
  \parprojl&: (\state_1, \state_2, \statevectorstar, \state_C)
  \mapsto
  (\state_1, \statevectorstar[C_1\mapsto C, C_2 \mapsto F], \state_C * \state_2)\\
  \parprojr&: (\state_1, \state_2, \statevectorstar, \state_C)
  \mapsto
  (\state_2, \statevectorstar[C_1 \mapsto F, C_2 \mapsto C], \state_C * \state_1)
\end{align*}
which give the state of the program from the point of view of one of the
programs in parallel.
This leads us to the following definition:
\begin{definition}\label{def:compatible}
  Two separated states $x_1,x_2\in\SStates$ are \emph{compatible} when
  there exists a three-party separated state $y$ such that $\parprojl(y) = x_1$ and
  $\parprojr(y) = x_2$. 
  Note that the three-party separated state $y$ is unique in that case.
\end{definition}

\paragraph{Framing}
To handle framing (\FRAME{} rule), we need to be able to move a piece of
(logical) heap from the Frame side to the Code side.
First, we define the framing of a logical state $\state_R$. Given an ATS $G$
over $\SepModelGraph$, we define $\framing{\state_R}(G)$ pointwise
with:
\[
  \getstate{}'(x) =
  \begin{cases*}
    (\state_C * \state_R, \statevector, \state_F) & if $\getstate{}(x) =
    (\state_C, \statevector, \state_F*\state_R)$\\
    \text{undefined} & otherwise
  \end{cases*}
\]
Given such a graph $G$ and a predicate $R$, we define $\framing{R}(G)$ as the
following union of ATSs (defined in~\S\ref{sec:cup-n-cap}):
\[
  \framing{R}(G) \quad = \quad \bigcup_{\state_R \vDash R} \framing{\state_R}(G)
\]
%

\paragraph{Resource introduction}
To give a semantics to the resource introduction rule, we need to extend the
hiding operator to separated states.
More precisely, the action of $\hide{r}$, in addition to replacing $P(r)$ and
$V(r)$ with $\knop$, is:
\begin{align*}
  (\state_C, \statevector \uplus [r\mapsto\state], \state_F) \quad&\mapsto\quad
  (\state_C * \state, \statevector, \state_F)\\
  (\state_C, \statevector \uplus [r\mapsto C \text{ or } F], \state_F) \quad&\mapsto\quad
  (\state_C, \statevector, \state_F)
\end{align*} 

\paragraph{Critical sections}
Similarly, we extend the definition of $\ATSwhen{r}(G)$ to separated states: to
the same underlying graph we associate the asynchronous morphism $\getstate{}'$
defined as
\[
  \getstate{}'(x) = (\state_C, \statevector \uplus [r\mapsto C], \state_F)
  \quad\text{ if }\; \getstate{G}(x) = (\state_C, \statevector, \state_F)
\]
We also need to take and release locks in the semantics of the proofs: the ATS
$\take{r}$ is defined by its Eve moves:
\begin{equation*}
  (\state_C, \statevector\uplus[r\mapsto\state], \state_F)
  \xrightarrow{\CODE{P(r)}}
  (\state_C * \state, \statevector\uplus[r\mapsto C], \state_F)
\end{equation*}
and $\release{r}$ by (for all $\state\in\LStates$ satisfying $r$'s "invariant" in
$\Gamma$):
\begin{equation*}
  (\state_C * \state, \statevector\uplus[r\mapsto C], \state_F)
  \xrightarrow{\CODE{V(r)}}
  (\state_C, \statevector\uplus[r\mapsto \state], \state_F).
\end{equation*}

\paragraph{Union}
It is sometimes necessary to combine in a single asynchronous strategy
two asynchronous strategies~$G_1$ and~$G_2$ whose purpose is to justify 
the transitions performed by the very same Code $C$.
This is precisely the purpose of the disjoint union~$G_1\cup G_2$.
The disjoint union $G_1 \cup G_2$ of two "ATSs" $G_1$ and $G_2$
over the model $\SepModelGraph$ of separated states
is defined in essentially the same way as the disjoint sum~$G_1 \oplus G_2$.
The only difference with the disjoint sum $G_1 \oplus G_2$ is that the
transitions (or moves) in $G_1 \cup G_2$ are meant to justify the transitions of
the \emph{same} program $C$, whereas the transitions (or moves) of the disjoint
sum $G_1 \oplus G_2$ are meant to justify the transitions of two different
programs~$C_1$ and~$C_2$.
So, given two derivation trees in CSL
\[
\prfsummary[\raiselabel{$\pi_1$}]{\prftree{\triple{\Gamma}{P_1}{C}{Q_1}}}
\quad\quad
\prfsummary[\raiselabel{$\pi_2$}]{\prftree{\triple{\Gamma}{P_2}{C}{Q_2}}}
\]
the asynchronous homomorphism associated to $\semSep{\pi_1} \cup \semSep{\pi_2}$ is
of the form
\[
  \semmorphsep_{\pi_1\cup\pi_2} \quad : \quad \semSep{\pi_1} \cup \semSep{\pi_2}
  \quad \longrightarrow \quad \semS{C}
\]
whereas the homomorphism associated to the sum is of the form:
\[
  \semmorphsep_{\pi_1\oplus\pi_2} \quad : \quad \semSep{\pi_1} \oplus \semSep{\pi_2}
  \quad \longrightarrow \quad \semS{C} \oplus \semS{C}
\]
The disjoint union~$\semmorphsep_{\pi_1\cup\pi_2}$ may be thus obtained by
postcomposing $ \semmorphsep_{\pi_1\oplus\pi_2}$ with the codiagonal
of~$\semS{C}$:
\[
\begin{array}{cl}
&
\begin{tikzcd}[column sep=2em]
\semSep{\pi_1} \oplus \semSep{\pi_2}\arrow[rr,"{\semmorphsep_{\pi_1\cup\pi_2}}"]
&&
\semS{C}
\end{tikzcd}
\\
= \!\!\!&
\begin{tikzcd}[column sep=2.2em]
\semSep{\pi_1} \oplus \semSep{\pi_2}\arrow[rr,"{\semmorphsep_{\pi_1\oplus\pi_2}}"]
&&
\semS{C}\oplus\semS{C}\arrow[rr,"codiagonal"]
&&
\semS{C}
\end{tikzcd}
\end{array}
\]

\paragraph{Intersection}
In order to interpret the introduction rule of the conjunction, we need to
define the intersection of two ATSs $G_1$ and $G_2$ over the same machine
model~$\genFPgraph$.
The definition of $G_1\cap G_2$ is very similar to the definition of the
"parallel product".
The only difference lies in the fact that the two ATSs $G_1$ and $G_2$ must
behave synchronously.
The nodes of $G_1 \cap G_2$ are defined in the same way as the nodes of
$G_1\|G_2$, that is, as the pairs $\interstate{x_1}{x_2}$ consisting of a
node~$x_1$ of $G_1$ and of a node~$x_2$ of $G_2$, such that
\[
\getstate{G_1}(x_1) \quad = \quad \getstate{G_2}(x_2)
\]
There are two types of transitions.
The Code transitions which are pairs of Code transitions:
\[
  \interstate{x_1}{x_2} \yrightarrow{\CODE{m}} \interstate{y_1}{y_2}
\]
when 
$\quad\;\;
  x_1 \yrightarrow{\CODE{m}} y_1 \;\in\; G_1
  \quad\text{ and }\quad
  x_2 \yrightarrow{\CODE{m}} y_2 \;\in\; G_2
$ \\
and similarly for Frame transitions.
A square is a "tile" in $G_1 \cap G_2$ precisely when
it is the superposition of two "tiles" of $G_1$ and of $G_2$.

\paragraph{Machine instructions}
The rules that correspond to machine instructions $m\in\Instr$ (such as $\LOAD$)
are interpreted in the obvious way, always preserving the permission associated
to affected locations.

\paragraph{Semantics of proofs}
\begin{align*}
  \Bigg\llbracket%
  \raisebox{-0.6em}{%
    \prftree%
    {\prfsummary[\raiselabel{$\pi_1$}]{\triple{\Gamma}{P}{C_1}{Q}}}%
    {\prfsummary[\raiselabel{$\pi_2$}]{\triple{\Gamma}{Q}{C_2}{R}}}%
    {\triple{\Gamma}{P}{C_1 ; C_2}{R}}  }%
  \Bigg\rrbracket_{\Sep}%
  \quad&=\quad%
  \semSep{\pi_1}\, ; \,\semSep{\pi_2}%
\intertext{%
For the parallel product rule \PAR, we use the "parallel product" of ATSs using
the above notion of \emph{compatibility}:}
  \Bigg\llbracket%
  \raisebox{-0.6em}{%
    \prftree%
    {\prfsummary[\raiselabel{$\pi_1$}]{\triple{\Gamma}{P_1}{C_1}{Q_1}}}%
    {\prfsummary[\raiselabel{$\pi_2$}]{\triple{\Gamma}{P_2}{C_2}{Q_2}}}%
    {\triple{\Gamma}{P_1*P_2}{C_1 \parallel C_2}{Q_1*Q_2}}   }%
  \Bigg\rrbracket_{\Sep}%
  \quad&=\quad%
  \semSep{\pi_1} \,\parallel\, \semSep{\pi_2}%
\\
  \Bigg\llbracket%
  \raisebox{-0.6em}{%
    \prftree%
    {\prfsummary[\raiselabel{$\pi$}]{\triple{\Gamma}{P}{C}{Q}}}%
    {\triple{\Gamma}{P*R}{C}{Q*R}}    }%
  \Bigg\rrbracket_{\Sep}%
  \quad&=\quad%
  \framing{R}\big(\sem{\pi}\big)%
\\
  \Bigg\llbracket%
  \raisebox{-0.6em}{%
    \prftree%
    {\prfsummary[\raiselabel{$\pi$}]{\triple{\Gamma, r:J}{P}{C}{Q}}}%
    {\triple{\Gamma}{P*J}{\resource{r}{C}}{Q*J}}   }%
  \Bigg\rrbracket_{\Sep}%
  \quad&=\quad%
  \hide{r}\big(\sem{\pi}\big)%
\\
%
  \Bigg\llbracket%
  \raisebox{-0.6em}{%
    \prftree%
    {\prfsummary[\raiselabel{$\pi$}]{\triple{\Gamma, r:J}{(P*J)\wedge B}{C}{Q*J}}}%
    {\triple{\Gamma, r:J}{P}{\when{r}{B}{C}}{Q}} }%
  \Bigg\rrbracket_{\Sep}%
  \quad&=\quad%
  \whentrue{B}\big(\take{r}\big)\,;\,\ATSwhen{r}\big(\semSep{\pi}\big) \,;\, \release{r}%
\\
  \Bigg\llbracket%
  \raisebox{-0.6em}{%
    \prftree%
    {\prfsummary[\raiselabel{$\pi_1$}]{\triple{\Gamma}{P_1}{C}{Q_1}}}%
    {\prfsummary[\raiselabel{$\pi_2$}]{\triple{\Gamma}{P_2}{C}{Q_2}}}%
    {\triple{\Gamma}{P_1 \vee P_2}{C}{Q_1 \vee Q_2}}    }%
  \Bigg\rrbracket_{\Sep}%
  \quad&=\quad%
  \semSep{\pi_1} \cup \semSep{\pi_2}  %
\intertext{The interpretation of the conjunction rule relies on the intersection of "ATSs"
just defined:}
  \Bigg\llbracket
  \raisebox{-0.6em}{
    \prftree
    {\prfsummary[\raiselabel{$\pi_1$}]{\triple{\Gamma}{P_1}{C}{Q_1}}}
    {\prfsummary[\raiselabel{$\pi_2$}]{\triple{\Gamma}{P_2}{C}{Q_2}}}
    {\triple{\Gamma}{P_1 \wedge P_2}{C}{Q_1 \wedge Q_2}}    }
  \Bigg\rrbracket_{\Sep}
  \quad&=\quad
  \semSep{\pi_1} \cap \semSep{\pi_2}
\end{align*}
Note that we do not rely on the preciseness of the context $\Gamma$ to define
the interpretation of the \CONJ{} proof rule, though "precision" is needed
in order to establish the soundness of the rule.

\section{An asynchronous soundness theorem}\label{section/soundness}
At this stage, we are ready to state our soundness theorem for Concurrent
Separation Logic.
%
%
%
We start by observing that every "proof" $\pi$ in CSL of a "Hoare triple" of the
form $\triplestd$ comes equipped with a "morphism of asynchronous graphs"
\[
  \semmorphsep_{\pi} \quad : \quad \semSep{\pi} \xrightarrow{\quad\quad} \semS{C}
\]
which makes the diagram below commute
\[
  \begin{tikzcd}[column sep = 6em, row sep=1.4em]
    \semSep{\pi} \arrow[r,"{\semmorphsep_{\pi}}"{description}] \arrow[d,"{\getstate{\pi}\,\,}"{swap}] & 
    \semS{C}
    \arrow[d,"{\,\,\getstate{C}}"]\\
    \SepModelGraph \arrow[r,"{\circledast}"{description}] & \StateModelGraph
  \end{tikzcd}
\]
The morphism~$\semmorphsep_{\pi}$ thus defines a morphism of ATS which relates
the interpretation of the proof~$\pi$ with the stateful interpretation of~$C$.
%
%
For every such "proof" $\pi$ of a "Hoare triple" $\triplestd$,
we write 
\[
    \semmorphfull_{\pi} \quad : \quad \semSep{\pi} \quad
    \xrightarrow{\quad\quad} \quad \semL{C}
\]
for the composite $\semmorphfull_{\pi}  =  \semmorph_C \circ \semmorphsep_\pi$ below:
\[
  \begin{tikzcd}[column sep=5em]
    \semSep{\pi} \arrow[r, "{\semmorphsep_{\pi}}"]& \semS{C} \arrow[r, "{\semmorph_C}"] &\semL{C}
  \end{tikzcd}
\]
%

\noindent
Our soundness theorem follows from two properties of the asynchronous strategy
$\semSep{\pi}$ associated to a CSL proof tree $\pi$. 
The first property (\emph{1-soundness}) implies that a well-specified program
does not crash during a \emph{valid execution}, that is, an execution which
starts from a state satisfying the precondition~$P$ and where the Frame performs
only legal transitions.
The second property (\emph{2-soundness}) implies that such a program does not
encounter any data race.
%
\begin{theorem}[1-soundness]\label{thm:1-soundness}
$\semmorphsep_{\pi}$ is a "Code $1$-fibration".
\end{theorem}
\noindent
A ""Code $1$-fibration"" is a "$1$-fibration" (Def.~\ref{def:1-fibration}) where we
only ask that Code transitions can be lifted, similarly to axiom~$\mathbf{3}$
of Def.~\ref{def:ATS-with-machine-model}.
This lifting property reflects the fact that the strategy~$\semSep{\pi}$
interpreting the proof~$\pi$ is \emph{winning}, in the sense that every
transition performed by the Code on machine states can be lifted (and thus
logically justified) by the strategy into a transition between separated states,
see \cite{mfps} for a discussion.
This implies in particular that \emph{well specified programs do not go wrong},
because the error state~$\Error$ cannot be lifted to a separated state.
The next statement is of a different nature: it says that the strategy
$\semSep{\pi}$ adapts at the \emph{separated} level to the possible reorderings
of scheduling performed at the \emph{stateless} level:
\begin{theorem}[2-soundness]\label{thm:2-soundness}
$\semmorphfull_{\pi}$ is a "$2$-fibration".
\end{theorem}
%

%
\noindent
This property implies (in particular) that valid executions of $C$ never produce
data races.
More deeply, it says that two executions which are equivalent modulo $\approx$
at the stateless level, in the sense that they behave in the same way with
respect to the locks (each thread acquires and releases each lock in the same
order), are also equivalent modulo~$\sim\,$ at the stateful level.
%
%
To make this statement formal, consider a well-specified program
$\triple{\emptyset}{P}{C}{Q}$ and define $\semSclosed{\{P\}C}{\tau}$ as the
subgraph of $\semS{C}$ obtained by removing every Frame transition, and keeping
only the states which can be reached from an "initial node" satisfying the
precondition $P$.
We are interested in the morphism
  \[
    \semmorph_{C}^{P} \quad : \quad \semSclosed{\{P\}C}{\tau} \quad \longrightarrow \quad \semL{C}
  \]
  obtained by restricting $\semmorph_C$ to the asynchronous subgraph
  $\semSclosed{\{P\}C}{\tau}$ of $\semS{C}$.
  This enables us to establish the soundness theorem announced in the
  introduction:
%
\begin{theorem}[Soundness]\label{thm/soundness}
$\semmorph_{C}^{P}$ is a "$2$-fibration".
\end{theorem}

\section{Conclusion and future works}\label{section/conclusion}
For the first time, we devise and establish a properly asynchronous version of
the Soundness Theorem for Concurrent Separation Logic (CSL).
%
%
%
%
%
In our formulation, the absence of data races follows from a more fundamental
lifting property of scheduling along the stateful-to-stateless translation
$\semS{C}\to\semL{C}$.
%
The proof of the theorem itself is original in design, and relies on the
construction of an asynchronous game semantics of CSL, building on the
foundations set in~\cite{mfps}.
In future work, we wish to adapt this asynchronous semantics of CSL
to weak memory models
\cite{boudol,brookes,Vafeiadis-separation-for-promising,promising-semantics,Dreyer-weak-memory-iris}
and to distributed algorithms \cite{Hongseok-strong-enough}.
In another direction of investigation, we want to extend our version of the
soundness theorem to a higher-order and axiomatic setting like Iris
\cite{iris1}.
Also, now that the asynchronous soundness theorem has been established by
semantic means, a nice and instructive challenge will be to prove it again using
purely syntactic techniques, in the line adopted for Mezzo~\cite{mezzo}.

\section*{Acknowledgments}
The authors are grateful to Richard Bornat, Stephen Brookes, Tony Hoare, Fran{\c
  c}ois Pottier and Viktor Vafeiadis for discussions at an early stage of this
work.

\bibliographystyle{alpha}
\bibliography{citations}

\newcommand{\etalchar}[1]{$^{#1}$}
\begin{thebibliography}{SPPD{\etalchar{+}}18}

\bibitem[BCOP05]{Bornat:permissions}
Richard Bornat, Cristiano Calcagno, Peter O'Hearn, and Matthew Parkinson.
\newblock Permission accounting in separation logic.
\newblock In {\em POPL}, 2005.

\bibitem[BCY06]{Bornat:sep}
Richard Bornat, Cristiano Calcagno, and Hongseok Yang.
\newblock Variables as resource in separation logic.
\newblock {\em ENTCS}, 155, 2006.

\bibitem[BP09]{boudol}
G{\'e}rard Boudol and Gustavo Petri.
\newblock Relaxed memory models: An operational approach.
\newblock POPL, 2009.

\bibitem[BPP14]{mezzo}
Thibaut Balabonski, Fran{\c{c}}ois Pottier, and Jonathan Protzenko.
\newblock Type soundness and race freedom for mezzo.
\newblock In {\em FLOPS}, 2014.

\bibitem[Bro04]{Brookes:a-semantics}
Stephen Brookes.
\newblock A semantics for concurrent separation logic.
\newblock In {\em CONCUR}, 2004.

\bibitem[FGH{\etalchar{+}}16]{dihomotopy-book}
Lisbeth Fajstrup, Eric Goubault, Emmanuel Haucourt, Samuel Mimram, and Martin
  Raussen.
\newblock {\em Directed Algebraic Topology and Concurrency}.
\newblock Springer, 2016.

\bibitem[GBC11]{Gotsman-precision}
Alexey Gotsman, Josh Berdine, and Byron Cook.
\newblock Precision and the conjunction rule in concurrent separation logic.
\newblock 2011.
\newblock MFPS.

\bibitem[GYF{\etalchar{+}}16]{Hongseok-strong-enough}
Alexey Gotsman, Hongseok Yang, Carla Ferreira, Mahsa Najafzadeh, and Marc
  Shapiro.
\newblock 'cause i'm strong enough: reasoning about consistency choices in
  distributed systems.
\newblock In {\em POPL}, 2016.

\bibitem[HW08]{hayman-Winskel}
Jonathan Hayman and Glynn Winskel.
\newblock Independence and concurrent separation logic.
\newblock {\em LMCS}, 2008.

\bibitem[JSS{\etalchar{+}}15]{iris1}
Ralf Jung, David Swasey, Filip Sieczkowski, Kasper Svendsen, Aaron Turon, Lars
  Birkedal, and Derek Dreyer.
\newblock Iris: Monoids and invariants as an orthogonal basis for concurrent
  reasoning.
\newblock In {\em POPL}, 2015.

\bibitem[KB17]{brookes}
Ryan Kavanagh and Stephen Brookes.
\newblock A denotational semantics for {SPARC} {TSO}.
\newblock MFPS, 2017.

\bibitem[KDD{\etalchar{+}}17]{Dreyer-weak-memory-iris}
Jan{-}Oliver Kaiser, Hoang{-}Hai Dang, Derek Dreyer, Ori Lahav, and Viktor
  Vafeiadis.
\newblock Strong logic for weak memory: Reasoning about release-acquire
  consistency in iris.
\newblock In {\em ECOOP}, 2017.

\bibitem[KHL{\etalchar{+}}17]{promising-semantics}
Jeehoon Kang, Chung{-}Kil Hur, Ori Lahav, Viktor Vafeiadis, and Derek Dreyer.
\newblock A promising semantics for relaxed-memory concurrency.
\newblock In {\em POPL}, 2017.

\bibitem[Mel17]{mellies-hdr}
Paul-Andr\'e Melli\`es.
\newblock {\em Une \'etude micrologique de la n\'egation}.
\newblock {HDR}, 2017.

\bibitem[MM07]{PAM:async-graph}
Paul{-}Andr{\'{e}} Melli{\`{e}}s and Samuel Mimram.
\newblock Asynchronous games: Innocence without alternation.
\newblock In {\em {CONCUR}}, 2007.

\bibitem[MS17]{mfps}
Paul-André Melliès and Léo Stefanesco.
\newblock A game semantics for concurrent separation logic.
\newblock In {\em MFPS}, 2017.

\bibitem[O'H07]{OHearn}
Peter~W. O'Hearn.
\newblock Resources, concurrency, and local reasoning.
\newblock {\em TCS}, 375, 2007.

\bibitem[PBC06]{Bornat:hoare}
Matthew~J. Parkinson, Richard Bornat, and Cristiano Calcagno.
\newblock Variables as resource in hoare logics.
\newblock In {\em LICS}, 2006.

\bibitem[Pra91]{HDA}
Vaughn Pratt.
\newblock Modeling concurrency with geometry.
\newblock POPL, 1991.

\bibitem[SPPD{\etalchar{+}}18]{Vafeiadis-separation-for-promising}
Kasper Svendsen, Jean Pichon-Pharabod, Marko Doko, Ori Lahav, and Viktor
  Vafeiadis.
\newblock A separation logic for a promising semantics.
\newblock 2018.
\newblock ESOP.

\bibitem[Vaf11]{Vafeiadis}
Viktor Vafeiadis.
\newblock Concurrent separation logic and operational semantics.
\newblock {\em ENTCS}, 276, 2011.

\end{thebibliography}

\newpage

\appendix

\section{Proof of the 1-soundness theorem (Thm.~\ref{thm:1-soundness})}
In this section we prove the 1-soundness theorem of CSL. The proof is done by
induction on the structure of the "proof tree".
Each case of the induction is given its own lemma.
We focus on the non-leaf rules of CSL.

\subsection{Parallel composition}
We begin with the rule for parallel composition. The corresponding case in the
induction is the following.
\begin{lemma}
Suppose that~$\pi$ is the "derivation tree"
  \[
    \prftree[r]{\PAR}
    {\prfsummary[\raiselabel{$\pi_1$}]{\triple{\Gamma}{P_1}{C_1}{Q_1}}}
    {\prfsummary[\raiselabel{$\pi_2$}]{\triple{\Gamma}{P_2}{C_2}{Q_2}}}
    {\triple{\Gamma}{P_1*P_2}{C_1 \parallel C_2}{Q_1*Q_2}}
  \]
and that the interpretation
\[
  \semmorphsep_i: \semSep{\pi_i} \to \semS{C_i}
\]
is a "$1$-fibration on Code transitions", for $i=1,2$.
In that case, the "asynchronous graph morphism" 
\[
  \semmorphsep: \semSep{\pi} \to\semS{C_1\parallel C_2}
\]
is also a "Code $1$-fibration".
\end{lemma}
A Code transition in $\semS{C_1 \parallel C_2}$ is (without loss of generality)
a pair of compatible transitions: one Code transition from $\semS{C_1}$ and one
Frame transition from $\semS{C_2}$.
We need to lift this transition into a move in $\semSep{\pi} = \semSep{\pi_1}
\parallel \semSep{\pi_2}$.
Using the induction hypothesis, we can show that Eve can lift the former into
$\semSep{\pi_1}$, and we can lift the latter because $\getstate{\pi_2}$ is a
Frame 1-fibration.
Therefore, we can lift the Code transition from $\semS{C_1\parallel C_2}$ into
$\semSep{\pi_1 \parallel \pi_2}$.

\begin{proof}
  Consider a node $x$ in $\semSep{\pi}$ whose label is $(\state_C, \statevector,
  \state_F)$, and let $\mstate = \bigsprod(\state_C, \statevector, \state_F)$.
  Consider a Code transition in $\semS{C_1\parallel C_2}$ of the form
  (writing the images of the nodes under$\getstate{}$)
  \begin{equation}\label{eq:transtition-parallel}
    \mstate \yrightarrow{\CODE{m}} \mstate'
  \end{equation}
  whose starting node is $\semmorphsep(x)$.
  By definition of the "parallel product" there exists a three-party separated
  state
  \[
    (\state_1, \state_2, \statevector, \state_F)
  \]
  whose projection through $\kparproj$ is $(\state_C, \statevector, \state_F)$.
  By definition of $\semS{C_1\parallel C_2}$, the
  transition~(\ref{eq:transtition-parallel}) is of the form
  \[
    a_1|a_2 \yrightarrow{\CODE{m}} b_1|b_2
  \]
  with $\getstate{1}(a_1) = \getstate{2}(a_2) = \mstate$ and $\getstate{1}(b_1) =
  \getstate{2}(b_2) = \mstate'$.
  Our goal is to find two moves, one in $\semSep{\pi_1}$ of the form
  \[
    (\state_1, \statevector, \state_F * \state_2) \yrightarrow{\CODE{m}}
    (\state'_1, \statevector', \state_F * \state_2)
  \]
  that is mapped under $\semmorphsep_1$ to the transition:
  \[a_1 \yrightarrow{m} a_2\] and one in $\semSep{\pi_2}$ of the form
  \[
    (\state_2, \statevector, \state_F * \state_1) \yrightarrow{\ENV{m}}
    (\state_2, \statevector', \state_F * \state'_1)
  \]
  that is mapped under $\semmorphsep_2$ to the transition:
\[b_1 \yrightarrow{\ENV{m}} b_2.\]
  (Note that we omit to write the change of perspective on the $\statevector$.)
  The first exists according to the hypothesis on $\semmorphsep_1$, and the
  second because $\getstate{2}$ is a "$1$-fibration" on Frame transitions.
\end{proof}

\subsection{Sequential composition}
The case of "sequential composition" is easy, since a Code transition of $C_1;
C_2$ is either a transition from $C_1$ or a transition from $C_2$, and both
cases follow immediately from the induction hypotheses.

\begin{lemma}
Suppose that~$\pi$ is the "derivation tree"
  \[
    \prftree[r]{\SEQ}
    {\prfsummary[\raiselabel{$\pi_1$}]{\triple{\Gamma}{P}{C_1}{Q}}}
    {\prfsummary[\raiselabel{$\pi_2$}]{\triple{\Gamma}{Q}{C_2}{R}}}
    {\triple{\Gamma}{P}{C_1 ; C_2}{R}}
  \]
and that the interpretation
\[
  \semmorphsep_i: \semSep{\pi_i} \to \semS{C_i}
\]
is a "$1$-fibration on Code transitions", for $i=1,2$.
In that case, the "asynchronous graph morphism"
\[
  \semmorphsep: \semSep{\pi} \to\semS{C_1; C_2}
\]
is also a "Code $1$-fibration".
\end{lemma}
\begin{proof}
  A Code transition in $\semS{C_1;C_2}$ is either a transition in $\semS{C_1}$
  or a transition in $\semS{C_2}$. The result follows from the hypotheses.
\end{proof}

\subsection{The frame rule}
\begin{lemma}
Suppose that~$\pi$ is the "derivation tree"
  \[
    \prftree[r]{\FRAME}
    {\prfsummary[\raiselabel{$\pi'$}]{\triple{\Gamma}{P}{C}{Q}}}
    {\triple{\Gamma}{P*R}{C}{Q*R}}
  \]
and that the interpretation
\[
  \semmorphsep: \semSep{\pi'} \to \semS{C}
\]
is a "$1$-fibration on Code transitions".
In that case, the "asynchronous graph morphism"
\[
  \semmorphsep: \semSep{\pi} \to\semS{C}
\]
is also a "Code $1$-fibration".
\end{lemma}
\begin{proof}
  Let us consider a node $x$ in $\semSep{\pi}$ and a Code transition
  \[
    a \yrightarrow{\CODE{m}} b \quad \in \quad \semS{C}
  \]
  where $\semmorphsep(x) = a$.
  Recall that the frame rule is interpreted as
  \[
    \framing{R}(\semSep{\pi'}) \quad = \quad \bigcup_{\state_R \vDash R}
    \framing{\state_R}(\semSep{\pi'})
  \]
  Therefore, the node $x$ belongs to one of the copies of
  $\framing{\state_R}(\semSep{\pi'})$, for some $\state_R$, and there is a node
  $x'$ above $a$ in $\semSep{\pi'}$ such that
  \[
    \getstate{}(x) = (\state_C * \state_R, \statevector, \state_F) \;\;\text { and }\;\;
    \getstate{}'(x') = (\state_C, \statevector, \state_F * \state_R)
  \]
  with $\state_R \vDash R$.
  By the hypothesis on $\semmorphsep'$, there is a node $y' \in \semSep{\pi'}$ and a
  move
  \[
    x' \yrightarrow{\CODE{m}} y'
  \]
  above the Code transition above.
  Hence, $\getstate{}'(y')$ is of the form
  \[
    \getstate{}'(y') \quad=\quad (\state_C', \statevector', \state_F*\state_R)
  \]
  which implies that there is a transition
  \[
    x \yrightarrow{\CODE{m}} y
  \]
  above the Code transition in $\framing{\state_R}(\semSep{\pi'}) \subseteq
  \semSep{\pi}$ with
  \[
    \getstate{}(y) \quad=\quad (\state_C' * \state_R, \statevector', \state_F)
  \]
\end{proof}

\subsection{Resource introduction}

The rule for "resource" introduction is interpreted using the $\hide{r}$
construction, which hides the new resource.
The proof consists basically in showing that if some Code transition $t$ can be
lifted into a move $T$, then $\hide{r}(t)$ can be lifted intro $\hide{r}(T)$.

\begin{lemma}
  Suppose that $\pi$ is the "derivation tree"
  \[
    \prftree[r]{\RES}
    {\prfsummary[\raiselabel{$\pi'$}]{\triple{\Gamma, r:J}{P}{C'}{Q}}}
    {\triple{\Gamma}{P*J}{\resource{r}{C'}}{Q*J}}
  \]
  and that the interpretation
  \[
    \semmorphsep': \semSep{\pi'} \to \semS{C'}
  \]
  is a "$1$-fibration on Code transitions".
  In that case, the "asynchronous morphism"
  \[
    \semmorphsep\quad:\quad \semSep{\pi} \;\longrightarrow\; \semS{\resource{r}{C'}}
  \]
  is also a "$1$-fibration on Code transitions".
\end{lemma}
\begin{proof}
  Let us consider a node $x$ in $\semSep{\pi}$ and a Code transition~$t$
  \[
    a \yrightarrow{\CODE{m}} b
  \]
  where $\semmorphsep(x) = a$.
  By definition of $\semS{\resource{r}{C'}}$, $t$ is the image under $\hide{r}$
  of a transition $t'$:
  \[
    a \yrightarrow{\CODE{\underline{m}}} b \quad \in \quad \semS{C'}
  \]
  \paragraph{Case 1.} Suppose $\underline{m}$ is neither $P(r)$ nor $V(r)$. Then
  $m=\underline{m}$ and the situation is:
  \[
    \getstate{}(x) = (\state_C * \state, \statevector, \state_F) \;\;\text { and }\;\;
    \getstate{}'(x) = (\state_C, \statevector\uplus[r\mapsto \state], \state_F)
  \]
  By the hypothesis on $\semmorphsep'$, there is a node $y \in \semSep{\pi'}$ and a
  move
  \[
    x \yrightarrow{\CODE{m}} y
  \]
  above the Code transition $t'$ and such that
  \[
    \getstate{}'(y) = (\state'_C, \statevector'\uplus[r\mapsto \state], \state_F)
  \]
  Hence, there is a lifting of $t$ starting from $x$.

  \paragraph{Case 2.} Suppose, for example, that $\underline{m} = P(r)$. In that case,
  $m = \knop$, and
  \[
    \getstate{}(x) = (\state_C * \state, \statevector, \state_F) \;\;\text { and }\;\;
    \getstate{}'(x) = (\state_C, \statevector\uplus[r\mapsto \state], \state_F)
  \]
  By the hypothesis on $\semmorphsep'$, there is a node $y \in \semSep{\pi'}$ and a
  move
  \[
    x \yrightarrow{\CODE{m}} y
  \]
  above the Code transition $T'$ and such that
  \[
    \getstate{}'(y) = (\state'_C * \state, \statevector'\uplus[r\mapsto C], \state_F)
  \]
  This means that there is a move above $T$ of the form (with labels instead of
  nodes):
  \[
    (\state_C * \state, \statevector, \state_F) \yrightarrow{\CODE{\knop}}
    (\state_C * \state, \statevector, \state_F)
  \]
\end{proof}

\subsection{Critical sections}
\begin{lemma}
  Suppose that $\pi$ is the "derivation tree"
  \[
    \prftree[r]{\WHEN}
    {\prfsummary[\raiselabel{$\pi'$}]{\triple{\Gamma}{(P*J)\wedge B}{C'}{Q*J}}}
    {\triple{\Gamma, r:J}{P}{\when{r}{B}{C'}}{Q}}
  \]
  and that the interpretation
  \[
    \semmorphsep'\quad:\quad \semSep{\pi'} \longrightarrow \semS{C'}
  \]
  is a "$1$-fibration on Code transitions". In that case, the "asynchronous
    morphism"
  \[
    \semmorphsep\quad:\quad \semSep{\pi} \longrightarrow \semS{\when{r}{B}{C'}}
  \]
  is also a "$1$-fibration on Code transitions".
\end{lemma}
\begin{proof}
  Recall that, in that case, the semantics of the Code and of the "derivation
    trees" are defined as
  \begin{align*}
    \semSep{C} &=
                 \whentrue{B}(\semS{P(r)} ; \ATSwhen{r}(\semS{C'}) ;\semS{V(r)})\\ & \qquad \juxt \whenabort{B}\\
    \semSep{\pi} &= \whentrue{B}(\take{r});\ATSwhen{r}(\semSep{\pi'}) ; \release{r}
  \end{align*}
  Consider a Code transition $t$ of $\semS{C}$
  \[
    a \yrightarrow{\CODE{m}} b
  \]
  and a node $x$ in $\semSep{C}$ above a.
  According to the side-condition $P \Rightarrow \defpred(B)$, we know that this
  transition is not in $\whenabort{B}$, because all the variables in $B$ are
  necessarily well-defined.
  There are now three possibilities:
  \paragraph{Case 1:} $t\in\ATSwhen{r}(\semS{C'}).$
  In that case the image of $x$ under $\getstate{}$ is of the form
  \[
    \getstate{}(x) = (\state_C, \statevector\uplus[r\mapsto C], \state_F)
  \]
  and $t$ is also a transition in $\semSep{C'}$, whose image under
  $\getstate{}'$ is:
  \[
    (\memstate, L \uplus \{r\}) \yrightarrow{\CODE{m}} (\memstate', L' \uplus \{r\})
  \]
  such that the image of $t$ under $\getstate{}$ is:
  \[
    (\memstate, L) \yrightarrow{\CODE{m}} (\memstate', L').
  \]
  Moreover, by hypothesis, we can lift the transition $t$ from $\semS{C'}$ to a
  move whose labels are of the form
  \[
    (\state_C, \statevector, \state_F) \yrightarrow{\CODE{m}}
    (\state'_C, \statevector', \state_F)
  \]
  Therefore, the same move is a lifting of $t$ in $\semSep{C}$, and its labels
  are
  \[
    (\state_C, \statevector\uplus[r\mapsto C], \state_F) \yrightarrow{\CODE{m}}
    (\state'_C, \statevector'\uplus[r\mapsto C], \state_F)
  \]
  \paragraph{Cases 2 and 3:} $t$ is in $\whentrue{B}(\semS{P(r)})$ or in
  $\semS{V(r)}$.
  Follows from the definitions.
\end{proof}

\subsection{The conjunction}

In this section, we suppose that $\Gamma$ is \emph{"precise"} and that $\pi$ is
the following "derivation tree":
\[
  \prftree[r]{\CONJ}
  {\prfsummary[\raiselabel{$\pi_1$}]{\triple{\Gamma}{P_1}{C}{Q_1}}}
  {\prfsummary[\raiselabel{$\pi_2$}]{\triple{\Gamma}{P_2}{C}{Q_2}}}
  {\triple{\Gamma}{P_1 \wedge P_2}{C}{Q_1 \wedge Q_2}}
\]

This rule is not sound when the context $\Gamma$ is not "precise",
see~\cite{Gotsman-precision}. In out setting, "precision" implies that Eve has at
most one way of lifting a given transition from the Code.
This is very useful, because, on the one hand, the induction hypothesis tells us
that there are two ways of lifting the same Code transition.
And, on the other hand, "precision" implies that they are actually the same, and
therefore define a move in the interpretation of \CONJ, which is defined using
a synchronous product.

\begin{lemma}
  Suppose the interpretation
  \[
    \semmorphsep_i: \semSep{\pi_i} \to \semS{C}
  \]
  is a "$1$-fibration on Code transitions", for $i=1,2$.
  In that case, the "asynchronous graph morphism" 
  \[
    \semmorphsep: \semSep{\pi} \to\semS{C}
  \]
  is also a "Code $1$-fibration".
\end{lemma}
\begin{proof}
  Consider a Code transition of $\semS{C}$
  \[
    a \yrightarrow{\CODE{m}} b
  \]
  and a node $x$ in $\semSep{\pi}$ above a.
  By definition of the semantics of $\pi$, there exist two nodes
  $x_1\in\semSep{\pi_1}$ and $x_2 \in \semSep{\pi_2}$ each above $a$ and such that
  \[
    \getstate{1}(x_1) = \getstate{2}(x_2) = \getstate{}(x).
  \]
  Therefore, according to the hypotheses there exist two moves in $\semSep{\pi_1}$
  and in $\semSep{\pi_2}$ each of the form
  \[
    x_i \yrightarrow{\CODE{m}} y_i.
  \]
  It suffices to show that $\getstate{}(y_1) = \getstate{}(y_2)$, since it
  implies that there exists $y$ in $\semSep{\pi}$ such that
  \[
    x \yrightarrow{\CODE{m}} y
  \]
  is above the transition on $\semS{C}$ we considered.

  Let us show, then, that $\getstate{}(y_1) = \getstate{}(y_2)$. Since all the
  permissions are preserved, by definition of the strategies, the first and the
  last component of the two "separated states" coincide. Moreover, "precision" tells
  us that there is at most one sub-logical state that satisfies the "invariants",
  which means that the middle components coincide as well the middle
  components coincide as well.
\end{proof}

\section{Proof of the 2-soundness theorem (Thm.~\ref{thm:2-soundness})}
In this section, we establish the 2-soundness theorem (Theorem~\ref{thm:2-soundness})
by induction on the "proof" $\pi$ of a "Hoare triple" of the form $\triplestd$.
First, we begin with the case of the $\PAR$ rule, which we find the most
interesting.


\subsection{The parallel rule}
\begin{lemma}
Suppose that~$\pi$ is the "derivation tree"
  \[
    \prftree[r]{\PAR}
    {\prfsummary[\raiselabel{$\pi_1$}]{\triple{\Gamma}{P_1}{C_1}{Q_1}}}
    {\prfsummary[\raiselabel{$\pi_2$}]{\triple{\Gamma}{P_2}{C_2}{Q_2}}}
    {\triple{\Gamma}{P_1*P_2}{C_1 \parallel C_2}{Q_1*Q_2}}
  \]
and that the interpretation
\[
\semmorphfull_i: \semSep{\pi_i} \to \semL{C_i}
\]
is a "2-fibration", for $i=1,2$.
In that case, the "asynchronous graph morphism" 
\[
\semmorphfull: \semSep{\pi} \to\semL{C_1\parallel C_2}
\]
is also a "2-fibration".
\end{lemma}
\begin{proof}
  Write $C = C_1 \parallel C_2$.
  Let us consider a "tile" in $\semL{C}$:
  \begin{equation}\label{diagram:2-par-instr-tile}
  \begin{tikzcd}[column sep=4em, row sep=\kTileHeight]
    & b_1|b_2 \ar[dr, "m'"]\\
    a_1|a_2 \ar[ur, "m"]\ar[dr, "m'"{swap}] & \approx & c_1|c_2\\
    & b'_1|b'_2 \ar[ur, "m"{swap}]
  \end{tikzcd}
\end{equation}
    such that there exist two transitions in $\semS{C}$
    \begin{equation}\label{diagram:2-par-sep-incomplete-tile}
  \begin{tikzcd}[column sep=4em, row sep=\kTileHeight]
    & y_1|y_2 \ar[dr, "m'"]\\
    x_1|x_2 \ar[ur, "m"] & & z_1|z_2
  \end{tikzcd}
\end{equation}
that are sent through $\semmorphfull$ onto the upper path
in~(\ref{diagram:2-par-instr-tile}).

    By definition of the "parallel product" of ATSs,
    since~(\ref{diagram:2-par-instr-tile}) is a "tile", its two projections are
    "tiles" as well:
    \[
      \begin{tikzcd}[column sep=1.5em, row sep=\kTileHeight]
        & b_1 \ar[dr, "{v'_{\restriction 1}}"]\\
        a_1 \ar[ur, "{u_{\restriction 1}}"]\ar[dr, "{v_{\restriction 1}}"{swap}] & \approx & c_1\\
        & b'_1 \ar[ur, "{u'_{\restriction 1}}"{swap}]
      \end{tikzcd}
      \quad\quad
      \begin{tikzcd}[column sep=2em, row sep=\kTileHeight]
        & b_2 \ar[dr, "{v'_{\restriction 2}}"]\\
        a_2 \ar[ur, "{u_{\restriction 2}}"]\ar[dr, "{v_{\restriction 2}}"{swap}] & \approx & c_2\\
        & b'_2 \ar[ur, "{u'_{\restriction 2}}"{swap}]
      \end{tikzcd}
    \]
    By hypothesis, the "asynchronous morphisms" $\semmorphfull_1$ and
    $\semmorphfull_2$ are both "2-fibrations". This means that the following two
    squares above them in~$\semS{C_1}$ and in~$\semS{C_2}$ are "tiles" as well:
    \[
      \begin{tikzcd}[column sep=1.5em, row sep=\kTileHeight]
        & y_1 \ar[dr, "{v'_{\restriction 1}}"]\\
        x_1 \ar[ur, "{u_{\restriction 1}}"]\ar[dr, "{s_1}"{swap}] & \sim & z_1\\
        & y'_1 \ar[ur, "t_1"{swap}]
      \end{tikzcd}
      \quad\quad
      \begin{tikzcd}[column sep=2em, row sep=\kTileHeight]
        & y_2 \ar[dr, "{v'_{\restriction 2}}"]\\
        x_2 \ar[ur, "{u_{\restriction 2}}"]\ar[dr, "s_2"{swap}] & \sim & z_2\\
        & y'_2 \ar[ur, "{t_2}"{swap}]
      \end{tikzcd}
    \]
    for some nodes $y_1'$ and $y'_2$ in $\semS{C_1}$ and $\semS{C_2}$
    respectively.
    By definition of "tiles" in $\semS{C_1 \parallel C_2}$, to show that there
    exists a "tile" completing~(\ref{diagram:2-par-sep-incomplete-tile})
    above~(\ref{diagram:2-par-instr-tile}), it suffices to show that the two
    states $\getstate{1}(y_1')$ and $\getstate{2}(y_2')$ are \emph{compatible},
    in the sense of Definition~\ref{def:compatible}.
    This follows from the following lemma.
\end{proof}

\begin{lemma}
  Suppose given two "ATSs" over "separated states" $G_1$ and $G_2$, and two "tiles":
    \[
      \begin{tikzcd}[column sep=1.5em, row sep=\kTileHeight]
        & y_1 \ar[dr, "{v'_{\restriction 1}}"]\\
        x_1 \ar[ur, "{u_{\restriction 1}}"]\ar[dr, "{s_1}"{swap}] & \sim & z_1\\
        & y'_1 \ar[ur, "t_1"{swap}]
      \end{tikzcd}
      \quad\quad
      \begin{tikzcd}[column sep=2em, row sep=\kTileHeight]
        & y_2 \ar[dr, "{v'_{\restriction 2}}"]\\
        x_2 \ar[ur, "{u_{\restriction 2}}"]\ar[dr, "s_2"{swap}] & \sim & z_2\\
        & y'_2 \ar[ur, "{t_2}"{swap}]
      \end{tikzcd}
    \]
    where the upper paths are compatible.
    Then $y'_1$ and $y'_2$, and the paths $s_1; t_1$ and $s_2; t_2$ define a
    path in $G_1 \parallel G_2$.
\end{lemma}
\begin{proof}
  By case analysis on the polarities of the transitions.
\end{proof}

\subsection{Sequential composition}
\begin{lemma}\label{lemma:2-soundness-sequential}
  Suppose that~$\pi$ is the "derivation tree"
  \[
    \prftree[r]{\SEQ}
    {\prfsummary[\raiselabel{$\pi_1$}]{\triple{\Gamma}{P}{C_1}{Q}}}
    {\prfsummary[\raiselabel{$\pi_2$}]{\triple{\Gamma}{Q}{C_2}{R}}}
    {\triple{\Gamma}{P}{C_1 ; C_2}{R}}
  \]
  and that the interpretation
  \[
    \semmorphfull_i: \semSep{\pi_i} \to \semL{C_i}
  \]
  is a "2-fibration", for $i=1,2$.
  In that case, the "asynchronous graph morphism"
  \[
    \semmorphfull: \semSep{\pi} \to\semL{C_1 ; C_2}
  \]
  is also a "2-fibration".
\end{lemma}
\begin{proof}
  Recall that the semantics of sequential composition is defined by:
  \begin{align*}
    \semL{C_1;C_2} &= \semL{C_1} ; \semL{C_2}\\
    \semSep{\pi_1;\pi_2} &= \semSep{\pi_1} \,;\, \semSep{\pi_2}
  \end{align*}
  This means that a "tile" in $\semL{C_1;C_2}$, is either a "tile" in $\semL{C_1}$
  or a "tile" in $\semL{C_2}$. By the hypothesis on $\semmorphfull_i$, it is clear
  that in either case we can lift the "tile" in either $\semSep{\pi_1}$ or in
  $\semSep{\pi_2}$, and thus in $\semSep{\pi_1 ; \pi_2}$.
\end{proof}

\subsection{Resource introduction}
\begin{lemma}
  Suppose that $\pi$ is the "derivation tree"
  \[
    \prftree[r]{\RES}
    {\prfsummary[\raiselabel{$\pi'$}]{\triple{\Gamma, r:J}{P}{C'}{Q}}}
    {\triple{\Gamma}{P*J}{\resource{r}{C'}}{Q*J}}
  \]
  and that the interpretation
  \[
    \semmorphfull': \semSep{\pi'} \to \semL{C'}
  \]
  is a "2-fibration".
  In that case, the "asynchronous morphism"
  \[
    \semmorphfull\quad:\quad \semSep{\pi} \;\longrightarrow\; \semL{\resource{r}{C'}}
  \]
  is also a "2-fibration".
\end{lemma}
\begin{proof}
  Write $C = \resource{r}{C'}$.
  Suppose there is a "tile" in $\semL{C}$ of the form
  \[
  \begin{tikzcd}[column sep=4em, row sep=\kTileHeight]
    & b \ar[dr, "m'"]\\
    a \ar[ur, "m"]\ar[dr, "m'"{swap}] & \approx & c\\
    & b' \ar[ur, "m"{swap}]
  \end{tikzcd}
  \]
  and that its upper path is the image of the following "tile" by $\semmorphfull$
  \begin{equation}\label{tile:path-resource-sep}
      \begin{tikzcd}
        & (\state'_C, \statevector_2, \state_F) \ar[dr, "m'"]\\
        (\state_C, \statevector_1, \state_F) \ar[ur, "m"] & &
        (\state''_C, \statevector_3, \state'_F)
    \end{tikzcd}
  \end{equation}
  Recall that the semantics of resource introduction is given by
  \begin{align*}
    \semL{\resource{r}{C'}} \quad &:= \quad \hide{r}(\semL{C'})\\
    \semSep{\RES(\pi')} \quad &:= \quad \hide{r}(\semSep{\pi'})
  \end{align*}
  By the definition of the semantics, it is the image under $\hide{r}$ of some "tile"
  \[
  \begin{tikzcd}[column sep=4em, row sep=\kTileHeight]
    & b \ar[dr, "\underline{m}'"]\\
    a \ar[ur, "\underline{m}"]\ar[dr, "\underline{m}'"{swap}] & \approx & c\\
    & b' \ar[ur, "\underline{m}"{swap}]
  \end{tikzcd}
  \]
  Suppose, first, that $\underline{m}$ touches $r$; say $\underline{m} = P(r)$ for
  example.
  Since the square above is a "tile", we know that, in that case, $\underline{m}'$
  is neither $P(r)$ nor $V(r)$.
  This implies that the path~(\ref{tile:path-resource-sep}) is the image by
  $\hide{r}$ of a path:
  \begin{equation*}
      \begin{tikzcd}[column sep=-2em]
        & (\underline{\state}_C*\state, \statevector \uplus [r \mapsto C], \state_F) \ar[dr, "m'"]\\
        (\underline{\state}_C, \statevector \uplus [r\mapsto\state], \state_F) \ar[ur, "P(r)"] & &
        (\underline{\state}''_C, \statevector_3\uplus[r\mapsto C], \state'_F)
    \end{tikzcd}
  \end{equation*}
  with $\state_C = \underline{\state}_C * \state$.
  By induction, this path can be completed into a "tile" in $\semSep{\pi'}$ of the
  form
  \begin{equation*}
      \begin{tikzcd}[column sep=-2em]
        & (\underline{\state}_C*\state, \statevector \uplus [r \mapsto C], \state_F) \ar[dr, "m'"]\\
        (\underline{\state}_C, \statevector \uplus [r\mapsto\state], \state_F) \ar[ur, "P(r)"] \ar[dr, "m'"'] & \sim &
        (\underline{\state}''_C, \statevector_3\uplus[r\mapsto C], \state'_F)\\
        & (\underline{\state}'''_C, \statevector \uplus [r \mapsto \state], \state_F)\ar[ur, "P(r)"']
    \end{tikzcd}
  \end{equation*}
  which implies that $\underline{\state}''_C = \underline{\state}'''_C * \state$.
  Finally, this implies that~(\ref{tile:path-resource-sep}) can be completed
  into a "tile":
  \begin{equation*}
      \begin{tikzcd}[column sep=0]
        & (\state_C, \statevector, \state_F) \ar[dr, "m'"]\\
        (\state_C, \statevector, \state_F) \ar[ur, "\knop"] \ar[dr, "m'"] & \sim &
        (\state''_C, \statevector_3, \state'_F)\\
        & (\state''_C, \statevector, \state'_F)\ar[ur, "\knop"]
    \end{tikzcd}
  \end{equation*}
  Note that this reasoning holds when $m'$ is replaced with a Frame move.
  (indeed, in the proof above, we accepted that $m'$ change $\state_F$).
  The case where neither $\underline{m}$ nor $\underline{m'}$ touch $r$ is
  similar.
\end{proof}

\subsection{Critical sections}
Before we prove the $2$-fibration lemma for the rule $\WHEN$, we analyze the
structure of $\ATSwhen{r}$.
Recall that this ATS is built in two steps (see
\S\ref{section/basic-operations}): first, the nodes and transitions of $G$ are
lifted by adding the new "resource" $r$ into the sets of locked "resources" of all
the states (and in the case of "separated states", we add that $r$ is locked by
the Code), and, second, we add Environment transitions to make it an ATS. Call
the first kind of transition \emph{""natural""}, and the second \emph{""artificial""}.
The "artificial" transitions correspond to the case where the Environment touches
the lock while it is held by the Code. This is, of course, a highly incorrect
behavior.
Thankfully, the constraints on the Frame moves in the "Machine model of Separated
States", and hence in the interpretation of proofs, rule these transitions out,
in the following sense.
First, "natural" transitions are stable by "homotopy".

\begin{lemma}
  Given a "tile" $T$ in $\ATSwhen{r}(G)$, if the upper path is made of two natural
  transitions, then so does the lower path.
\end{lemma}
\begin{proof}[Proof sketch]
  In a "tile", opposite transitions have the same "footprints", and therefore the
  same behavior on locks.
\end{proof}

\noindent
In the interpretations of proofs, all Code transitions are "natural".

\begin{lemma}
  If $G$ is an "ATS" over the "machine model of separated states", then all the Code
  transitions of $\ATSwhen{r}$ are "natural".
\end{lemma}
\begin{proof}[Proof sketch]
  Since the lock $r$ is held be the Code, the Frame cannot touch it.
\end{proof}

\begin{lemma}
  Suppose that $\pi$ is the derivation tree
  \[
    \prftree[r]{\WHEN}
    {\prfsummary[\raiselabel{$\pi'$}]{\triple{\Gamma}{(P*J)\wedge B}{C'}{Q*J}}}
    {\triple{\Gamma, r:J}{P}{\when{r}{B}{C'}}{Q}}
  \]
  and that the interpretation
  \[
    \semmorphfull'\quad:\quad \semSep{\pi'} \longrightarrow \semL{C'}
  \]
  is a 2-fibration. In that case, the asynchronous morphism
  \[
    \semmorphfull\quad:\quad \semSep{\pi} \longrightarrow \semL{\when{r}{B}{C'}}
  \]
  is also a 2-fibration.
\end{lemma}
\begin{proof}
  Write $C = \when{r}{B}{C'}$.
  Recall that the semantics of $C$ is defined as:
  \begin{multline*}
    \semSep{C'} = \whentrue{B}(\semSep{P(r)} ; \ATSwhen{r}(\semSep{C}) ;\semSep{V(r)})\\ \juxt \whenabort{B}.
  \end{multline*} 
  By definition of "sequential composition" of ATSs, a tile in $\semL{C}$ contains
  a Code transition from $\semL{P(r)}$ (or $\semL{V(r)}$) only if it is a
  Code/Frame tile.
  This case is easy because $P(r)$ and the Adam move touch distinct components
  of the separated states (since it is a tile, Adam cannot touch the $r$
  component of $\statevector$), and $\getstate{}$ is a Code-Frame $2$-fibration
  by definition of ATSs.

  Consider a path of the following form in $\semSep{\pi}$:
  \begin{equation}\label{path:critical-section}
    \begin{tikzcd}[column sep=-10pt]
      & (\state_C, \statevector_2\uplus[r\mapsto C], \state_F) \ar[dr, "m'"]\\
      (\state_C, \statevector\uplus[r\mapsto C], \state_F) \ar[ur, "m"] & &
      (\state''_C, \statevector_3\uplus[r\mapsto C], \state'_F)
    \end{tikzcd}
  \end{equation}
  and consider a tile in $\semL{C}$ whose upper path is the image of the above
  path under $\semmorphfull$.

  We can suppose that this tile is in $\ATSwhen{r}(G)$.
  According to the two lemmas above, it is of the form (where we write, instead
  of the nodes themselves, their images under $\getstate{}'$)
  \begin{equation}\label{tile:critical-section-C}
  \begin{tikzcd}[column sep=4em, row sep=\kTileHeight]
    &  L_2 \uplus \{r\} \ar[dr, "m'"]\\
    L_1 \uplus \{r\} \ar[ur, "m"]\ar[dr, "m'"{swap}] & \approx & L_3 \uplus \{r\}\\
    & L'_2 \uplus \{r\} \ar[ur, "m"{swap}]
  \end{tikzcd}
\end{equation}
  and, moreover, there is a tile in $\semL{C'}$ (on the same nodes, since the map
  $\ATSwhen{r}$ is defined pointwise) of the form:
  \begin{equation}
    \label{tile:critical-section-C-prime}
  \begin{tikzcd}[column sep=4em, row sep=\kTileHeight]
    &  L_2  \ar[dr, "m'"]\\
    L_1\ar[ur, "m"]\ar[dr, "m'"{swap}] & \approx & L_3 \\
    & L'_2 \ar[ur, "m"{swap}]
  \end{tikzcd}
  \end{equation} 
  By hypothesis, and by the definition of the semantics of the rule $\WHEN$,
  there exists a tile in $\semSep{\pi'}$ above the
  tile~(\ref{tile:critical-section-C-prime}), which is of the form:
  \[
    \begin{tikzcd}[column sep=0]
      & (\state_C, \statevector_2, \state_F) \ar[dr, "m'"]\\
      (\state_C, \statevector, \state_F) \ar[ur, "m"] \ar[dr, "m'"] & \sim &
      (\state''_C, \statevector_3, \state'_F)\\
      & (\state''_C, \statevector_2', \state'_F)\ar[ur, "m"]
    \end{tikzcd}
  \]
  where the domain of the $\statevector$'s does not contain $r$. This finally
  means that we can complete the path~(\ref{path:critical-section}) into the
  following "tile" above~(\ref{tile:critical-section-C}) in $\semSep{\pi}$:
  \[
    \begin{tikzcd}[column sep=-9pt]
      & (\state_C, \statevector_2\uplus[r\mapsto C], \state_F) \ar[dr, "m'"]\\
      (\state_C, \statevector\uplus[r\mapsto C], \state_F) \ar[ur, "m"] \ar[dr, "m'"] & \sim &
      (\state''_C, \statevector_3\uplus[r\mapsto C], \state'_F)\\
      & (\state''_C, \statevector_2'\uplus[r\mapsto C], \state'_F)\ar[ur, "m"]
    \end{tikzcd}
  \]
\end{proof}

\subsection{Conjunction}
\begin{lemma}
  Suppose that $\pi$ is the "derivation tree"
  \[
    \prftree[r]{\CONJ}
    {\prfsummary[\raiselabel{$\pi_1$}]{\triple{\Gamma}{P_1}{C}{Q_1}}}
    {\prfsummary[\raiselabel{$\pi_2$}]{\triple{\Gamma}{P_2}{C}{Q_2}}}
    {\triple{\Gamma}{P_1\wedge P_2}{C}{Q_1\wedge Q_2}}
  \]
  where $\Gamma$ is precise,
  and that the interpretation
  \[
    \semmorphfull_i\quad:\quad \semSep{\pi_i} \longrightarrow \semL{C}
  \]
  is a 2-fibration, for $i=1,2$. In that case, the "asynchronous morphism"
  \[
    \semmorphfull\quad:\quad \semSep{\pi} \longrightarrow \semL{C}
  \]
  is also a 2-fibration.
\end{lemma}
\begin{proof}
  Consider a "tile" $T$ in $\semL{C}$
  such that its upper path can be lifted to a path $p$ in $\semSep{\pi}$.
  %
  By definition of $\semSep{\pi} := \semSep{\pi_1} \cap \semSep{\pi_2}$, the
  path $p$ corresponds to a path $p_1$ in $\semSep{\pi_1}$ and $p_2$ in
  $\semSep{\pi_2}$, which all have the same image under $\getstate{}$,
  $\getstate{1}$ and $\getstate{2}$ respectively.
  Moreover, by hypothesis, the tile $T$ can be lifted to "tiles" $T_1$ and $T_2$
  in $\semSep{\pi_1}$ and in $\semSep{\pi_2}$ respectively, such that their
  upper paths are $p_1$ and $p_2$, respectively.
  %
  %
  Since there is at most one tile which has a given upper path (Axiom $2$ of
  asynchronous graphs), the two "tiles" have equal images under $\getstate{1}$ and
  $\getstate{2}$. Therefore $T_1$ and $T_2$ define a "tile" in $\semSep{\pi}$
  extending the path~$p$.
\end{proof}

\subsection{Conditionals}
\begin{lemma}
  Suppose $\pi$ is the following "derivation tree":
  \[
  \prftree[r]{\IF}
  {\prfsummary[\raiselabel{$\pi_1$}]{\triple{\Gamma}{P\wedge B}{C_1}{Q}}}
  {\prfsummary[\raiselabel{$\pi_2$}]{\triple{\Gamma}{P\wedge \neg B}{C_2}{Q}}}
  {\triple{\Gamma}{P}{C}{Q}}
  \]
  where $C := \ifte{B}{C_1}{C_2}$, and that the interpretation
  \[
    \semmorphfull_i\quad:\quad \semSep{\pi_i} \longrightarrow \semL{C_i}
  \]
  is a "2-fibration", for $i=1,2$. In that case, the "asynchronous morphism"
  \[
    \semmorphfull\quad:\quad \semSep{\pi} \longrightarrow \semL{C}
  \]
  is also a "2-fibration".
\end{lemma}
\begin{proof}
  Recall that the semantics of $\kif$ statements is defined as
\begin{align*}
  \semSep{\ifte{B}{C_1}{C_2}} &= \whentrue{B}(\semSep{\knop});\semSep{C_1}\\
                            &\;\juxt \whenfalse{B}(\semSep{\knop});\semSep{C_2}\\
                            &\;\juxt \whenabort{B}
\end{align*}
All non trivial "tiles" are either in $\semL{C_1}$ or in $\semL{C_2}$. In either
case, the hypothesis on $\semmorphfull_i$ tells us that this "tile" can be lifted
to $\semSep{\pi_i}$.
\end{proof}

\subsection{Disjunction}
\begin{lemma}
  Suppose that $\pi$ is the "derivation tree"
  \[
    \prftree[r]{\DISJ}
    {\prfsummary[\raiselabel{$\pi_1$}]{\triple{\Gamma}{P_1}{C}{Q_1}}}
    {\prfsummary[\raiselabel{$\pi_2$}]{\triple{\Gamma}{P_2}{C}{Q_2}}}
    {\triple{\Gamma}{P_1\vee P_2}{C}{Q_1\vee Q_2}}
  \]
  and that the interpretation
  \[
    \semmorphfull_i\quad:\quad \semSep{\pi_i} \longrightarrow \semL{C}
  \]
  is a "2-fibration", for $i=1,2$. In that case, the "asynchronous morphism"
  \[
    \semmorphfull\quad:\quad \semSep{\pi} \longrightarrow \semL{C}
  \]
  is also a "2-fibration".
\end{lemma}
\begin{proof}
  Similarly to the previous case, if $T$ is a tile in $\semL{C}$, there we can
  either lift it to $\semSep{\pi_1}$ or to $\semSep{\pi_2}$, so in any case we
  can lift it to $\semSep{\pi}$.
\end{proof}

\subsection{The frame rule}
\begin{lemma}
  Suppose that $\pi$ is the "derivation tree"
  \[
    \prftree[r]{\FRAME}
    {\prfsummary[\raiselabel{$\pi'$}]{\triple{\Gamma}{P}{C}{Q}}}
    {\triple{\Gamma}{P * R}{C}{Q * R}}
  \]
  and that the interpretation
  \[
    \semmorphfull'\quad:\quad \semSep{\pi'} \longrightarrow \semL{C}
  \]
  is a "2-fibration". In that case, the "asynchronous morphism"
  \[
    \semmorphfull\quad:\quad \semSep{\pi} \longrightarrow \semL{C}
  \]
  is also a "2-fibration".
\end{lemma}
\begin{proof}
  Consider a path in $\semSep{\pi}$ of the following form (it is in one of the
  $\framing{\state_R}(\semSep{\pi'})$)
  \begin{equation}\label{eq:path-frame-2fib}
    \begin{tikzcd}[column sep=15pt]
      & (\state'_C * \state_R, \statevector_2, \state_F) \ar[dr, "m'"]\\
      (\state_C * \state_R, \statevector_1, \state_F) \ar[ur, "m"] & &
      (\state''_C * \state_R, \statevector_3, \state_F)
    \end{tikzcd}
  \end{equation} 
  and a "tile" in $\semL{C}$ whose upper path is the image
  of~(\ref{eq:path-frame-2fib}) under $\semmorphfull_{\pi}$:
  \begin{equation}\label{eq:tile-framing}
  \begin{tikzcd}[column sep=7em, row sep=\kTileHeight]
    & b \ar[dr, "m'"]\\
    a \ar[ur, "m"]\ar[dr, "m'"{swap}] & \approx & c\\
    & b' \ar[ur, "m"{swap}]
  \end{tikzcd}
  \end{equation} 
  By definition of $\framing{\state_R}$, and according to the hypothesis on
  $\semmorphfull_{\pi'}$, the path~(\ref{eq:path-frame-2fib}) in
  $\framing{\state_R}(\semSep{\pi'})$ corresponds to a path in $\semSep{\pi'}$
  that is the upper path a "tile" of the form:
  \begin{equation*}
    \begin{tikzcd}[column sep=15pt]
      & (\state'_C, \statevector_2, \state_F * \state_R) \ar[dr, "m'"]\\
      (\state_C, \statevector_1, \state_F * \state_R) \ar[ur, "m"] \ar[dr, "m'"]& \sim &
      (\state_C, \statevector_3, \state_F * \state_R) \\
       &(\state^\dagger_C, \statevector'_3, \state_F * \state_R) \ar[ur, "m"]
    \end{tikzcd}
  \end{equation*} 
  which means that there is a "tile" that extends~(\ref{eq:path-frame-2fib})
  above~(\ref{eq:tile-framing}):
  \begin{equation*}
    \begin{tikzcd}[column sep=15pt]
      & (\state'_C * \state_R, \statevector_2, \state_F) \ar[dr, "m'"]\\
      (\state_C * \state_R, \statevector_1, \state_F) \ar[ur, "m"] \ar[dr, "m'"]& \sim &
      (\state_C * \state_R, \statevector_3, \state_F) \\
       &(\state^\dagger_C * \state_R, \statevector'_3, \state_F) \ar[ur, "m"]
    \end{tikzcd}
  \end{equation*} 

\end{proof}

\end{document}

%
\[
\begin{tikzcd}[column sep=1.5em]
% ...
\]
%
\[
\begin{tikzcd}[column sep=1.5em]
\]
\[
%
\]
%
\[
%
\]
%
\[
%
\]
